\documentclass{aa}  

\usepackage{bm} 
\usepackage{mathrsfs}
\usepackage{graphicx}
\usepackage{txfonts}
\usepackage[colorlinks=true,citecolor=blue,linkcolor=blue,]{hyperref}
\usepackage{multirow}
\usepackage{mathrsfs}

\newcommand{\Teff}{T_{\rm eff}}
\newcommand{\Eq}[1]{Eq.~(\ref{#1})}
\newcommand{\Eqs}[2]{Eqs.~(\ref{#1}) to (\ref{#2})}

\newcommand{\cP}{c_{\rm P}}
\newcommand{\cV}{c_{\rm V}}
\newcommand{\Sec}[1]{Sect.~\ref{#1}}
\newcommand{\xxx}{\bm{x}}
\newcommand{\Ekin}{E_{\rm kin}}

\begin{document} 

\title{Magnetohydrodynamic simulations of A-type stars:\\ Long-term evolution of core dynamo cycles}


\author{J.P. Hidalgo\inst{1} \and P. J. K\"apyl\"a\inst{2} \and D. R. G Schleicher\inst{1}  \and C. A. Ortiz-Rodríguez\inst{3} \and F. H. Navarrete\inst{4,5}} 

\institute{Departamento de Astronomía, Facultad Ciencias Físicas y Matemáticas, Universidad de Concepción, Av. Esteban Iturra s/n Barrio Universitario, Concepción, Chile. \and
Institut f\"ur Sonnenphysik (KIS), Georges-K\"ohler-Allee 401a, 79110 Freiburg, Germany. \and
Hamburger Sternwarte, Universit\"at Hamburg, Gojenbergsweg 112, 21029
Hamburg, Germany. \and 
Institute of Space Sciences (ICE-CSIC), Campus UAB, Carrer de Can Magrans s/n, 08193, Barcelona, Spain. \and
Institut d’Estudis Espacials de Catalunya (IEEC), Carrer Gran Capità 2–4, 08034 Barcelona, Spain.}

   \date{Received XXX; accepted XXX}

 
  \abstract
{Early-type stars have convective cores due to a steep temperature
 gradient produced by the CNO cycle. These cores can host dynamos,
  and the generated magnetic fields can be relevant to
  explain the magnetism observed in Ap/Bp stars.}
{Our main objective is to characterise the convective core dynamos and
  differential rotation, and to do the first quantitative analysis of
  the relation
  between magnetic activity cycle and rotation period.}
{We use numerical 3D star-in-a-box simulations of a $2.2~M_\odot$ A-type
  star with a convective core of roughly 20\% of the stellar radius
  surrounded by a radiative envelope. Rotation rates from 8 to 20 days
  are explored. We use two
  models of the entire star, and an additional zoom set, where 50\% of
  the radius is retained.}
{The simulations produce hemispheric core dynamos with cycles and
  typical magnetic field strengths around
  $60$~kG. However, only a very small fraction of the magnetic
  energy is able to reach the surface. The cores have solar-like
  differential rotation, and a substantial part of the radiative
  envelope has
  quasi-rigid rotation. In the most rapidly rotating cases the
  magnetic energy in the core
  is roughly 40\% of the kinetic
  energy. Finally, we find that the magnetic cycle period
  $P_\mathrm{cyc}$ increases with decreasing the rotation period
  $P_\mathrm{rot}$ which is also observed in many simulations of
  solar-type stars.}
{Our simulations indicate that a strong hemispherical core dynamo
  arises routinely, but that it is not enough the explain
  the surface magnetism of Ap/Bp stars. Nevertheless, as the core
    dynamo produces dynamically relevant magnetic fields it should not
    be neglected when other mechanisms are explored.}


   \maketitle
%

\section{Introduction}

Magnetic fields can be found in a wide variety of stars, and there is
a general consensus that most of them are amplified and maintained via
self-excited dynamos. In this context, these processes typically
require rotation and fluid motions, and therefore are most likely to
occur inside convection zones (see \citealt{Brandenburg-2005, 2017LRSP...14....4B}). The
most obvious example is the Sun, which has a cyclic large-scale
magnetic field driven by a self-excited dynamo operating in its
convective envelope (see e.g. \citealt{Charbonneau-2020}). Like the
Sun, other late-type stars (M, K, F) have convective envelopes, and
the observed magnetic fields in these stars have also very likely a
convective origin. Numerical simulations targeting a wide variety of
rotating fully or partially convective stars yield dynamos of various
sorts \citep[see, e.g.][and references
  therein]{2023SSRv..219...58K}.

Early-type main-sequence stars have masses $M > 1.5~M_\odot$ and
effective temperatures $\Teff \gtrsim 10^4$~K. Roughly $10\%$ of
early-type stars are magnetic \citep[see e.g.][]{Kochukhov-2006, Landstreet-2007,Landstreet-2008, Grunhut-2017, 2019MNRAS.490..274S}. According to our
understanding of stellar evolution, these stars have radiative, and
therefore stably stratified, envelopes. This might explain why
a dynamo cannot usually operate near the surfaces of these stars,
since vigorous fluid motions are absent. However, some
stars in the range $1.5~M_\odot$ to $6~M_\odot$ host observable
large-scale magnetic fields. The best example is the subgroup of
chemically peculiar main-sequence stars, classified as Ap/Bp. These
stars host magnetic fields with mean field strengths ranging from
$200~\mathrm{G}$ to $30~\mathrm{kG}$ \citep{Auriere-2007}, the
strongest field being $34~\mathrm{kG}$ in HD 215441
\citep{Babcock-1960}. Roughly $10\%$ of A-type stars
belong to the sub-classification Ap and have detectable strong
magnetic fields \citep{Moss-2001}. The non-magnetised population
(“normal” A-type stars) usually have very weak fields below the
detection limit. Nevertheless, weak magnetic fields have been detected
in the normal population, for example in Sirius and Vega. In the case
of Vega, Zeeman polarimetry gives a magnetic field strength of
$0.6\pm0.3$~G \citep{Lignieres-2009, Petit-2010, Petit-2022}, and
$0.2\pm0.1$~G for Sirius \citep{Petit-2011}. Normal A stars are
typically rapid
rotators \citep{Royer-2007}, whereas Ap stars are slower rotators than
their non-magnetic counterparts \citep{Abt1995}. Indeed,
\cite{Mathys2008} have shown that non-magnetic A stars have rotation
periods ranging from a few hours to a day, while most Ap stars have
periods between one and ten days.
This could be an indication of magnetic braking,
although there is no clear correlation between rotation and magnetic
fields \citep{Kochukhov-2006}.

The origin of the large-scale magnetic fields in Ap/Bp stars remains
uncertain, although several theories have been proposed. The fossil
field theory suggests that these magnetic fields originate from the
protostellar phase of the
star. The Ohmic diffusion time in radiative zones is very long
(e.g. $10^{10}$ years in the context of the Sun,
\citealt{Cowling-1945}), and therefore, a magnetic field in
equilibrium in a radiative zone could survive the whole main-sequence
life of the star. Stable magnetic configurations in fully radiative A-type stars have been reported in numerical simulations
\citep{2004Natur.431..819B, Braithwaite-2006, Braithwaite-2008, Becerra-2022}. In
principle, the fossil magnetic field has to survive the convective
protostellar phase as the star descends the Hayashi track
\citep{Hayashi-1962, Siess-2000}. This appears to be very unlikely
assuming a fully convective evolution, but a transition to a radiative
phase could counterbalance this problem and preserve a significant 
part of the initial magnetic flux \citep{Schleicher-2023}. 
Such protostellar models were proposed e.g. by \cite{Palla1992, Palla1993}.

Alternatively, it has been proposed that radiative envelopes can
host dynamos, as a result of the interaction between a magnetic
instability \citep[e.g.][]{Tayler-1973} and 
differential rotation. This is the Tayler-Spruit dynamo scenario
\citep{Spruit-2002}. Recently,
\cite{Petitdemange-2023,Petitdemange-2024} performed global
simulations of a magnetised stably stratified fluid with differential
rotation in a spherical shell, reporting the first numerical
demonstration
of the Tayler-Spruit dynamo.

Another alternative is a strong core dynamo. Early-type
stars have convective cores due to the temperature sensitivity of the
dominant nuclear reaction (CNO cycle) such that energy production is
highly concentrated in the centre of the star, leading to a steep
temperature gradient that drives convection. It was
long suspected that the convective cores in A stars host dynamos
\citep{Krause-1976}. Simulations by \cite{Browning-2004} showed that
such convective cores have differential rotation that is beneficial
for dynamo action. This was reported
by \cite{Brun-2005}, who performed numerical simulations of the inner
$30\%$ by radius (half of which was convective) of a
$2~M_\odot$ A-type
star, obtaining magnetic fields with typical strengths around 
equipartition with kinetic energy. 
Interestingly, the inclusion of a fossil field in the radiative 
envelope might affect 
the nature of the core dynamo \citep[see, e.g.][in the
context of the Sun]{Boyer-1984}. 
\cite{Featherstone-2009} performed simulations of
the core dynamo from a $2~M_\odot$ A-type star, reporting an 
equipartition magnetic field. However, the inclusion of a twisted
toroidal fossil field can lead to a superequipartition state in 
the core, where the magnetic energy is roughly 10 times
stronger than the kinetic energy.
In B-type stars, dynamos
have been found as well. Simulations by \cite{Augustson-2016}
of the core of a $10~M_\odot$ B-type star, showed vigorous dynamo
action, and generated superequipartition magnetic fields with peak
values exceeding a megagauss in the rapid rotators. However,
although these studies show that a convective core can lead to strong
dynamo action, the nature of these dynamos has not been explored in
detail. For example, no cyclic solutions or a relation between the
magnetic cycle period to rotation period of the core dynamo have been
reported. 

A natural question is whether the magnetic fields generated in the core can
reach the surface
of the star. This could happen under the action of
buoyancy. Stellar structure models predict convection zones
  close to the surface of early-type
  stars \citep{Richard-2001, Cantiello-2009} due to bumps in the
  stellar opacities from iron \citep{Iglesias-1992}, hydrogen and
  helium ionisation. These layers can host dynamos and
  the resulting magnetic field, could
  easily rise to the surface due to magnetic buoyancy, creating surface
  magnetic fields of order a few Gauss
  \citep{Cantiello-2019}.  
  In principle, the same could happen in the context of a core
  dynamo. However, the timescale of this process has 
  been
estimated to be longer than the main-sequence lifetime of these stars
\citep{Schuessler-1978, Parker-1979, Moss-1989}, unless the magnetic
structures are very small. This was reported by \cite{MacGregor-2003},
who modelled a buoyant magnetic flux ring in the radiative interior of an
early-type star (with $M < 10~M_\odot$), and found transport
timescales that were shorter than the main-sequence lifetime of the
star. Nonetheless, \cite{MacDonald-2004} re-examined
the conclusions of \cite{MacGregor-2003} with more
realistic models including
differential
rotation and realistic strong compositional gradients in the
radiative layers. This slows down the buoyancy process considerably,
and magnetic fields higher than the equipartition values are required
to make this process feasible. One mechanism that could counter this
situation is convective overshooting. This could allow mixing in the
stably stratified zone above the convective core, reducing the
compositional gradient. Also, the field that penetrates the radiative
envelope might be strengthened by the shearing produced by the
differential rotation, making it strong enough to reach the surface
due to buoyancy. 

Core dynamo simulations often include the convective core and only a
part of the radiative envelope of the star \citep[e.g.][]{Brun-2005,
  Augustson-2016}. In the current study, we perform simulations of a
main-sequence A-type star using the star-in-a-box model. This setup
allows us to explore dynamo action generated by convection in the core
of the star (including $r=0$), and study the resulting magnetic field
not only in the core but also on the stellar surface, modelling the
entire star for the first time in 3D numerical simulations. Another
point of interest is to study magnetic cycles and their relation to
the rotation period. The relation between these quantities has been
studied before
in simulations of other types of stars \citep[see e.g.][for solar-like
  stars]{Warnecke-2018, Strugarek-2018}, but such relation has not
been presented before in the context of A-type stars. The models as
well as the methods are described in Section~\ref{Methods}. The
analysis and results of the simulations are provided in
Section~\ref{results}. Discussion and conclusions of this study
are presented in Section~\ref{conclusions}.

\section{Numerical models and methods \label{Methods}}
\subsection{Full star setup \label{full-star}}
The model used here is based on the star-in-a-box setup presented by
\cite{Kapyla-2021}, which is based on the setup by
\cite{Dobler-2006}. The computational domain corresponds to a box of
side $l=2.2R$, where $R$ is the stellar radius and all coordinates
$(x,y,z)$ range from $-l/2$ to $l/2$. The following non-ideal fully
compressible MHD equation set is solved:
\begin{align}
    \frac{\partial \bm{A}}{\partial t} &= \bm{U} \times \bm{B} - \eta \mu_0 \bm{J}, \label{s3mhd_1} \\
    \frac{D \ln \rho}{D t} &= - \bm{\nabla} \bm{\cdot} \bm{U}, \label{s3mhd_2} \\
    \frac{D \bm{U}}{D t} &= - \bm{\nabla} \Phi - \frac{1}{\rho} \left( \bm{\nabla} p - \bm{\nabla} \bm{\cdot}2 \nu \rho \bm{\mathsf{S}} + \bm{J} \times \bm{B} \right)- 2 \bm{\Omega}\times \bm{U} + \bm{f}_d, \label{s3mhd_3} \\
    T \frac{Ds}{Dt} &= - \frac{1}{\rho} \left[ \bm{\nabla} \bm{\cdot} (\bm{F}_\text{rad} + \bm{F}_\text{SGS})  + \mathcal{H} - \mathcal{C} + \mu_0 \eta \bm{J}^2 \right] + 2 \nu \bm{\mathsf{S}}^2, \label{s3mhd_4}
\end{align}
where $\bm{A}$ is the magnetic vector potential, $\bm{U}$ is the flow
velocity, $\bm{B} = \bm{\nabla} \times \bm{A}$ is the magnetic field,
$\eta$ is the magnetic diffusivity, $\mu_0$ is the magnetic
permeability of vacuum, $\bm{J} = \bm{\nabla} \times \bm{B}/\mu_0$ is
the current density given by Ampère's law, $D/Dt = \partial/\partial t
+ \bm{U} \bm{\cdot}\bm{\nabla}$ is the advective (or material)
derivative, $\rho$ is the mass density, $p$ is the pressure, $\Phi$ is
the fixed gravitational potential, given by 
the Pad\'e approximation obtained from 1D stellar structure model:
\begin{equation}
  \Phi(r') = - \frac{GM}{R} \frac{a_0 + a_2 r'^2 + a_3 r'^3}{1 + b_2r'^2 + b_3 r'^3 + a_3 r'^4}, \label{grav-a0}
\end{equation}
where $r'=r/R$ is the fractional radius, $G$ is the gravitational
constant, $M$ is the mass of
the star. For an A0-type star the coefficients are given by $a_0 =
4.3641$, $a_2 = -1.5612$, $a_3 = 0.4841$, $b_2 = 4.0678$, and $b_3 =
1.2548$. $\nu$ is the kinematic viscosity, $\bm{\mathsf{S}}$ is the traceless
rate-of-strain tensor, given by
\begin{equation}
  \mathsf{S}_{ij} = \frac{1}{2}(\partial_j U_i + \partial_i U_j) - \frac{1}{3}\delta_{ij} \bm{\nabla} \bm{\cdot} \bm{U}, \label{S-tensor}
\end{equation}
where $\delta_{ij}$ is the Kronecker
delta. $\bm{\Omega}=(0,0,\Omega_0)$ is the rotation vector and
$\bm{f}_\mathrm{d}$ describes damping of flows
exterior to the star, given by
\begin{equation}
    \bm{f}_d = - \frac{\bm{U}}{\tau_{\text{damp}}} f_e(r), \label{damping}
\end{equation}
where $\tau_{\text{damp}} = 0.2\tau_\mathrm{ff} \approx 
1.5~\mathrm{days}$ is the damping timescale and $\tau_\mathrm{ff} =
\sqrt{R^3/GM}$ is the freefall time.
The function $f_e(r)$ is defined as
\begin{equation}
    f_e(r) = \frac{1}{2} \left(1 + \tanh\frac{r - r_{\text{damp}}}{w_{\text{damp}}} \right), \label{f_e-damping}
\end{equation}
where $r_{\text{damp}}= 1.03R$ is the damping radius and
$w_{\text{damp}}=0.02R$ is its width. $T$ is the temperature, $s$ is
the
specific entropy, $\bm{F}_\mathrm{rad}$ is the radiative flux and
$\bm{F}_\mathrm{SGS}$ is the subgrid-scale (SGS) entropy
flux. Radiation
inside the star is approximated as a diffusion process. Therefore, the
radiative flux is given by
\begin{align}
    \bm{F}_\mathrm{rad} = - K \bm{\nabla} T, \label{radiative-flux}
\end{align}
where $K$ is the radiative heat conductivity, a quantity that is
assumed to be constant. In addition, it is convenient to introduce a
SGS entropy diffusion that does not contribute to the
net energy transport, but damps fluctuations near the grid scale. This
is given by the SGS entropy flux
\begin{equation}
    \bm{F}_\mathrm{SGS} = -\chi_\mathrm{SGS} \rho \bm{\nabla}s', \label{entropy-flux}
\end{equation}
where $\chi_\mathrm{SGS}$ is the SGS diffusion coefficient, and
\begin{equation}
    s' = s - \langle s \rangle_t
\end{equation}
is the fluctuating entropy, where $\langle s \rangle_t(\xxx)$ is a
running temporal mean of the specific entropy. $\mathcal{H}$ and
$\mathcal{C}$ describe additional heating and cooling, respectively,
and we adopted similar expressions as in \cite{Dobler-2006} and
\cite{Kapyla-2021}, where
\begin{equation}
    \mathcal{H}(r) = \frac{L_{\text{sim}}}{(2\pi w_L^2)^{3/2}} \exp \left( - \frac{r^2}{2 w_L^2} \right),
\end{equation}
is a normalised Gaussian profile that parameterises the
nuclear energy production inside the core of the star, where
$L_{\text{sim}}$ is the luminosity in the simulation, and $w_L = 0.1R$
is the width of the Gaussian. Furthermore, $\mathcal{C}(\bm{x})$ models the
radiative losses above the stellar surface, and it is given by
\begin{equation}
    \mathcal{C}(\bm{x}) = \rho c_\text{P} \frac{T(\bm{x}) - T_{\text{surf}}}{\tau_{\text{cool}}}f_e(r),
\end{equation}
where $\cP$ is the heat capacity at constant pressure,
$T_{\text{surf}}=T(R)$ is the temperature at the surface of the star,
$\tau_{\text{cool}} = \tau_{\text{damp}}$ is a cooling timescale, and
$f_e(r)$ is given by \Eq{f_e-damping}, with the same
parameters, $r_{\text{cool}} = r_{\text{damp}}$ and $w_{\text{cool}} =
w_{\text{damp}}$. The ideal gas equation of state $p = \mathcal{R}\rho
T$ is assumed, where $\mathcal{R}=\cP-\cV$ is the ideal gas constant, 
and $\cV$ is the heat capacity at constant volume.

To avoid diffusive spreading of magnetic fields and flows from the
core to the envelope we follow \cite{Kapyla-2202}, and use
radial profiles for the diffusivities $\nu$ and $\eta$, where the
radiative layers have values $10^2$ times smaller than the core.
The magnetic diffusivity has to be already rather low in the core to
excite a dynamo; see \Sec{results}. Therefore the diffusivities in the
radiative envelope are even lower making it hard to resolve the flows
and magnetic fields there. To cope with this issue, we add
artificial sixth-order hyperdiffusivity terms in the dynamical 
equations. The hyperdiffusive terms smooth small scale oscillation
to avoid numerical instabilities near grid scale
\cite[see e.g.][]{Brandenburg-2002-V2, Johansen-2005}. Here we use the
resolution-independent mesh hyper-Reynolds number method described in
the Appendix A of \cite{Lyra-2017} for $\nu$ and $\eta$, in all the
sets of simulations.
Furthermore, the advective terms of \Eqs{s3mhd_1}{s3mhd_4} are written
in terms of fifth-order upwinding derivatives with a sixth-order
hyperdiffusive correction with a flow dependent diffusion coefficient;
see Appendix B of \cite{Dobler-2006}.

\subsection{Zoom setup \label{zoom-setup}}

Modelling the whole star allows us to study the convective core all
the way to the surface of the star. However, this also reduces the
number of
grid points available to resolve convection inside the core. To test
the robustness of the results and to increase the spatial resolution
in the core, we run a few simulations using a zoom model, which is
otherwise the
same as the full star model, but the box has a side of
$l=1.1R$. Therefore, we exclude the surface of the star and focus
exclusively in the convective core with higher resolution.

The flow damping radius was changed to $r_\mathrm{damp} =
0.45 R$ with $\tau_\mathrm{damp}=0.02\tau_\mathrm{ff} \approx 0.15$ days, 
and
$w_\mathrm{damp} = 0.01 R$, and the cooling now starts at
$r_\mathrm{cool} = 0.52 R$, with $w_\mathrm{cool}=0.02R$ and
$\tau_\mathrm{cool} = 0.2\tau_\mathrm{ff}\approx 1.5~\mathrm{days}$.
These choices were made more out of numerical convenience rather than
physical arguments. However, as will be shown in \Sec{results} the
main results remain unaffected.

\subsection{Physical units, initial and boundary conditions}

The stellar parameters are extracted from a $2.2~M_\odot$ main-sequence
one dimensional stellar model. This model was obtained using the MESA
code \citep[see][and the references therein]{Paxton-2019}. The radius,
mass density and temperature of the stellar centre, and luminosity
averaged from $t = 2.5 \cdot 10^6$ years to $t = 4.5 \cdot 10^8$ years
are given by $R_* = 2.1 ~ R_\odot$, $\rho_{0} = 5.5 \cdot
10^4~\mathrm{kg}\,\mathrm{m^{-3}}$, $T_{0} = 2.3 \cdot
10^{7}~\mathrm{K}$ and $L_* = 23.5 ~ L_\odot$, where $L_\odot =
  3.83\cdot 10^{26}$~W is the solar luminosity,
respectively. Furthermore, the surface gravity is:
\begin{equation}
    g_* = \frac{G M_*}{R_*^2} \approx \frac{2.2}{2.1^2} \frac{G M_\odot}{R_\odot^2} \approx 0.5 g_\odot =  137~\frac{\mathrm{m}}{\mathrm{s^2}}, \label{surface-gravity}
\end{equation}
where $g_\odot = 274~\mathrm{m\,s^{-2}}$.
Using realistic parameters in fully compressible simulations of
stars is infeasible due
to the huge gap between the acoustic (dynamic) and
Kelvin-Helmholtz (thermal) timescales. A solution to
this problem was originally pointed out by \cite{Chan-1986}, and it
consists of exaggerating the luminosity in numerical simulations. Here
we follow the notation of \cite{Dobler-2006} and define a
dimensionless luminosity
\begin{eqnarray}
\mathcal{L} = \frac{L}{\sqrt{G^3 M^5/R^5}},
\end{eqnarray}
with which we define the luminosity ratio between the simulation
  and the target star:
\begin{eqnarray}
L_\mathrm{ratio} = \frac{\mathcal{L}_\mathrm{sim}}{\mathcal{L}_\mathrm{star}}.
\end{eqnarray}
Now, the gap
between acoustic and Kelvin-Helmholtz timescales is reduced by a
cumulative factor of 
$L_\mathrm{ratio}^{4/3}$ \citep[see][for a detailed
  discussion]{Kapyla-2021}.

The units of length and time are given by the radius of the star $[x]
= R$ and the free-fall time $[t] = \tau_\mathrm{ff}$
respectively. Furthermore, the unit of magnetic field is obtained from
the equipartition field strength as $[B] = \sqrt{\mu_0 \rho_0} [x]/[t]$,
and the unit of entropy is $[s] = \cP$. The conversion factors between
simulation and physical units are \citep[for example, $x = x_\mathrm{fac}
x_\mathrm{sim}$, see][]{Kapyla-2020}
\begin{gather}
    x_\mathrm{fac} = \frac{R_*}{R_\mathrm{sim}}, \hspace*{0.5cm}  t_\mathrm{fac} = \frac{\Omega_\mathrm{sim}}{\Omega_*}, \hspace*{0.5cm} U_\mathrm{fac} = \frac{R_*\Omega_*}{R_\mathrm{sim} \Omega_\mathrm{sim}}, \\
    \rho_\mathrm{fac} = \frac{\rho_0}{\rho_\mathrm{0,sim}},\hspace*{0.5cm}  B_\mathrm{fac} = \left[ \frac{\mu_0 \rho_0 (\Omega_* R_*)^2}{\mu_\mathrm{0,sim} \rho_\mathrm{0,sim} (\Omega_\mathrm{sim} R_\mathrm{sim})^2}\right]^{1/2},
\end{gather}
where the subscript “sim” represents the quantity in simulation
units. Furthermore, since the convective velocity scales with luminosity as
$u_\mathrm{conv} \propto L^{1/3}$
\citep{Jones-2017, Kapyla-2020, Baraffe-2023, 2024A&A...683A.221K},
the rotation rate needs to be enhanced by the same factor
($L_\mathrm{ratio}^{1/3}$) to have a consistent rotational influence
on the flow. Following Appendix A of \cite{Kapyla-2020}, we obtain
\begin{equation}
    \Omega_\mathrm{sim} = L_\mathrm{ratio}^{1/3} \left( \frac{g_\mathrm{sim}}{g_*} \frac{R_*}{R_\mathrm{sim}} \right)^{1/2} \Omega_*.
\end{equation}
The convective core is assumed to encompass $20\%$ of the stellar
radius. To set such configuration we assume a piecewise polytropic
initial state. A polytrope is
defined by
\begin{equation}
    p(\rho) = K_0 \rho^{\gamma},
\end{equation}
where $K_0$ is a constant, and
\begin{equation}
    \gamma = \frac{d \ln p}{d \ln \rho} = 1 + \frac{1}{n} \label{gamma}
\end{equation}
is the adiabatic index written in terms of the polytropic index
$n$. We choose $n=n_{\rm ad}=1.5$ in the convectively unstable layer
($r < 0.2R$), and $n=n_{\rm rad}=3.25$ in the stable layer ($r >
0.2R$) with a smooth
transition between them. Here $n_{\rm ad}$ corresponds to a marginally
stable stratification whereas $n_{\rm rad}$ arises in a hydrostatic
solution of a radiative atmosphere with Kramers opacity law
\citep[e.g.][]{2014A&A...571A..68B}.

The boundary conditions are impenetrable and stress-free conditions
for flow and the magnetic field is assumed to be perpendicular to the
boundary. We further
assume a vanishing second derivative for $\ln \rho$, and vanishing
temperature gradient across the exterior boundaries of the box. For
the initial conditions of the flow and magnetic field, we considered
low amplitude Gaussian noise with initial amplitudes of 
$2\cdot10^{-3}$~m/s and $1$~G respectively.

The simulations were run with the {\sc Pencil
  Code}\footnote{\url{https://pencil-code.org/}}
\citep{Brandenburg-2002, Pencil-code-2021}, which is a highly modular
high-order finite-difference code for solving partial differential
equations.

\subsection{Dimensionless parameters}

To characterise our simulations, the following dimensionless numbers
are computed. The effect of rotation relative to
viscous forces is measured by the Taylor number
\begin{equation}
    \mathrm{Ta} = \frac{4 \Omega_0^2 \Delta r^4}{\nu^2},
\end{equation}
where $\Delta r = 0.2R$ is the depth of the convective zone.
The magnetic and SGS Prandtl numbers are
\begin{align*}
    \mathrm{Pr}_\mathrm{M} = \frac{\nu}{\eta}, \hspace*{0.5cm} \mathrm{Pr}_\mathrm{SGS} = \frac{\nu}{\chi_\mathrm{SGS}}.
\end{align*}
The influence of rotation on the
flow is measured by the Coriolis number
\begin{equation}
    \mathrm{Co} = \frac{2 \Omega_0}{u_\mathrm{rms} k_R},
\end{equation}
where $u_\mathrm{rms}$ is the root-mean-square (rms) velocity averaged over
the convective zone ($r < \Delta r$) and $k_R = 2\pi/\Delta r$ is the
scale of the largest convective eddies. 
The fluid and magnetic Reynolds numbers are defined as
\begin{align}
    \mathrm{Re} = \frac{u_\mathrm{rms}}{\nu k_R}, \hspace*{0.5cm}  \mathrm{Re_M} = \frac{u_\mathrm{rms}}{\eta k_R},
\end{align}
and the SGS P\'eclet number is
\begin{equation}
    \mathrm{Pe} = \frac{u_\mathrm{rms}}{\chi_\mathrm{SGS} k_R}.
\end{equation}
Furthermore, the Brunt-Väisälä (or buoyancy) frequency and the 
Richardson number related to rotation in the radiative zone are defined 
as
\begin{equation}
    N = \sqrt{\frac{g}{c_\mathrm{P}} \frac{ds}{dr}},\hspace*{0.5cm} \mathrm{Ri}_\Omega = \frac{N^2}{\Omega_0^2},
\end{equation}
respectively. If $N^2 > 0$, then the layer will be stable 
against buoyancy fluctuations.

\section{Results \label{results}}

\begin{table*}[h!]
\centering
\caption{Summary of the simulations.}
\begin{tabular}{lccccccccccc}
\hline\hline\noalign{\smallskip}
Run & $P_\mathrm{rot}$~[days] & $u_\mathrm{rms}$~[m/s] & $B_\mathrm{rms} $~[kG]  &
Co & Ta[$ 10^{8}$] & Pe & Re & $\mathrm{Re}_\mathrm{M}$ & $\mathrm{Ri}_\Omega[10^{-2}]$ & $P_\mathrm{cyc}$~[years]\\
\hline\noalign{\smallskip}
MHDr1 & 20 & 99 & no dynamo & 5.0 & 1.10 & 11 & 52 & 36 & 16.0 & -  \\
MHDr2 & 15 & 51 & 60 & 10.1 & 1.95 & 7 & 35 & 24 & 9.2 & $1.81  \pm  0.10$ \\
MHDr2* & 15 & 52 & 57 & 9.9 & 1.95 & 7 & 35 & 25 & 9.2 & $2.14  \pm  0.13$  \\
MHDr3 & 10 & 39 & 65 & 19.2 & 4.38 & 5 & 27 & 19 & 4.1 & $2.50  \pm  0.11$ \\
MHDr3* & 10 & 39 & 62 & 18.8 & 4.38  & 5 & 28 & 19 & 3.9 & $2.62  \pm  0.14$ \\
MHDr4 & 8 & 35 & 60 &  26.9 & 6.84 & 5 & 24 & 17 & 2.6 & $3.14  \pm  0.05$ \\
MHDr4* & 8 & 35 & 57 & 26.2 & 6.84 & 5 & 25 & 17 & 2.5 & $3.37  \pm  0.08$  \\
\hline
zMHDr1 & 20 & 95 & not saturated  & 4.9 & 1.10 & 11 & 53 & 37 & 16.8 & -  \\
zMHDr2 & 15 & 55 & 49 &  9.6 & 1.95 & 7 & 36 & 25 &  9.6 & $1.95  \pm  0.03$  \\
zMHDr3 & 10 & 39 & 53 &  18.9 & 4.38 & 5 & 28 & 19 & 4.3 & $2.60  \pm  0.03$ \\
zMHDr4 & 8 & 36 & 46 &  26.9 & 6.84 & 5 & 24 & 17 & 2.7 & $2.50  \pm  0.02$ \\
\hline
\end{tabular}
\tablefoot{From left to right the columns
  correspond to the rotation period $P_\mathrm{rot} = 2\pi/\Omega_0$,
  the volume averaged (over the convective zone) rms flow
  velocity $ u_\mathrm{rms}$, the volume-averaged rms
  magnetic field $B_\mathrm{rms}$, the Coriolis number, the Taylor
  number, the SGS P\'eclet number, the fluid and magnetic Reynolds
  numbers, the Richardson number averaged over the radiative zone, and
  the magnetic cycle period.}
\label{table-1}
\end{table*}

We present three sets of simulations exploring rotation periods from 8
to 20 days. All simulations have $\nu =
5.45\cdot 10^7~\mathrm{m^2/s}$, $\eta = 7.78\cdot
10^7~\mathrm{m^2/s}$, and $\chi_\mathrm{SGS} = 2.61\cdot
10^8~\mathrm{m^2/s}$ in the core, and therefore
$\mathrm{Pr}_\mathrm{M} = 0.7$ and $\mathrm{Pr}_\mathrm{SGS} \approx
0.21$. In the first set of simulations (group MHD), $\nu$ and
$\eta$ have radial jumps around $0.35R$, with a smooth transition over
a width of $0.06R$. Above the jump, diffusivities are $10^2$ times
smaller. In the second set (group MHD*) the jump is at $0.3R$ and
its width is $0.03R$. Simulations in this group start from already
saturated snapshots of the respective simulation with the same
rotation rate in the MHD group. The third set
corresponds to the zoom setup (group zMHD) described in
Section~\ref{zoom-setup}. The radial profiles are the same as those
in group MHD. The runs, as well as the diagnostic quantities are
listed in Table~\ref{table-1}. All simulations were run on a grid of
$200^3$ uniformly distributed grid points.

\begin{figure}[t!]
    \centering
    \includegraphics[width=\hsize]{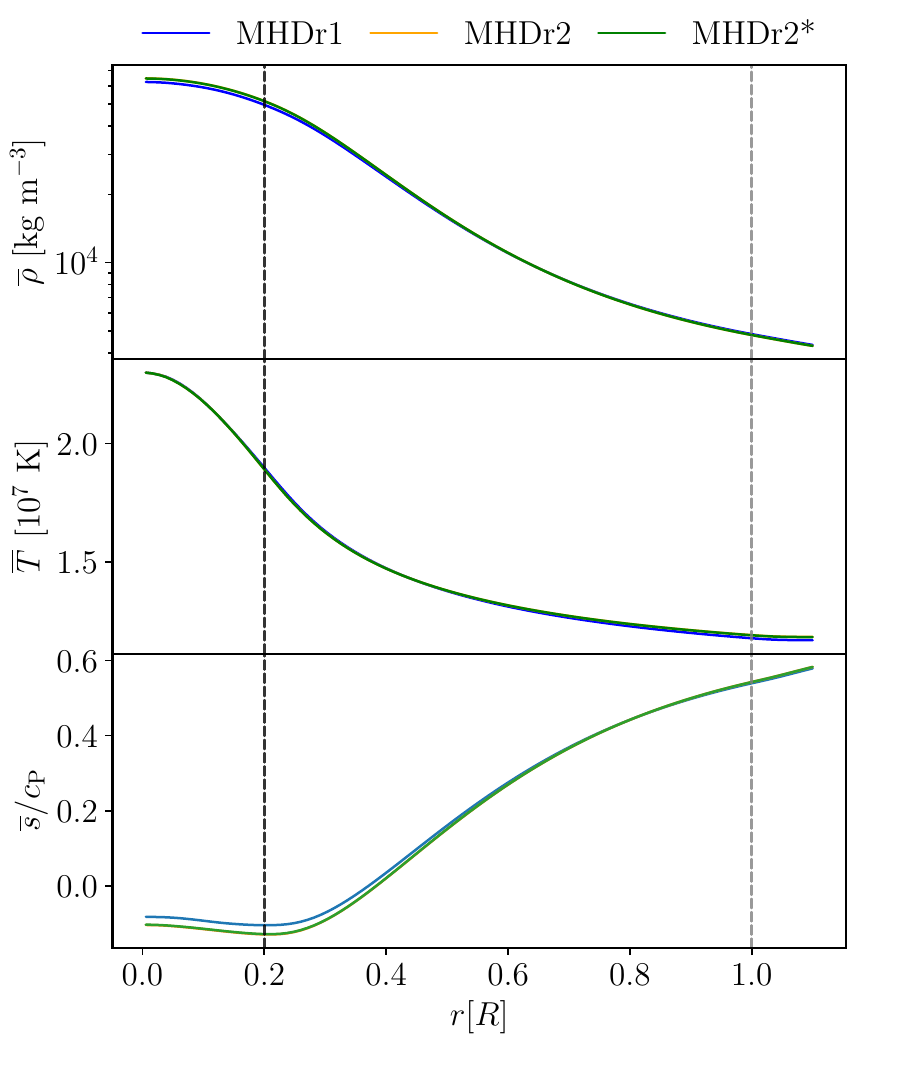}
    \caption{\textit{Top panel}: Density profiles as functions of
      radius from MHDr1, 
      MHDr2, and
      MHDr2*. \textit{Middle panel:} Temperature profiles for the same runs. \textit{Bottom panel}:
      Specific entropy profiles from the same runs. The black
        dashed line indicates the surface of the convective core ($r=0.2R$) and the grey dashed line the stellar surface ($r=R$). The
      quantities are time-averaged over the thermally relaxed phase.}
    \label{fig:strat}
\end{figure}

Figure~\ref{fig:strat} shows the radial profiles of density, temperature, and
specific entropy from representative runs, where the overlines denote 
horizontal ($\phi \theta$) averaging.
The density stratification between the centre and the surface of the star ($r=R$) is
$\overline{\rho}_\mathrm{centre}/\overline{\rho}_\mathrm{surface}
\approx 13$ in the full star runs. Furthermore, the stratification
between the centre and the surface of the convective core ($r=0.2R$) is
$\overline{\rho}_\mathrm{centre}/\overline{\rho}_\mathrm{surfcore}
\approx 1.27$, which is close to that from the MESA model
$\overline{\rho}_\mathrm{centre}/\overline{\rho}_\mathrm{surfcore}
\approx 1.5$. The same ratios for the temperature are
  $\overline{T}_\mathrm{centre}/\overline{T}_\mathrm{surface} \approx
  2$ and $\overline{T}_\mathrm{centre}/\overline{T}_\mathrm{surfcore}
  \approx 1.23$. From the lower panel of Figure~\ref{fig:strat},
we can infer a
negative entropy gradient in the convective core and a positive
gradient at the rest of the star, which is the expected
configuration for a A-type main-sequence star. The surface
  cooling becomes effective above $r_{\rm cool}$ and relaxes the
  regions exterior to the star toward a constant temperature $T_{\rm
    surf}$. Therefore the density drops nearly exponentially above
  $r=r_{\rm cool}$. Radiation transports all of the energy in the
  stellar envelope all the way to the surface; see
  Sect.~\ref{sec:fluxes}. This means that the photosphere at the
  stellar surface is modelled in a very simplistic way where the energy
  transport mechanism switches from radiative diffusion to the cooling
  flux which mimics radiative losses in the optically thin
  exterior. This approach avoids detailed physics leading to, for
  example, thin convection zones near the surface. Since the focus in
  the current study is in the core dynamo, this simplification is
  justified.

Finally, the
stratification in the cores of the simulations in group zMHD is
similar to those in the full star models, and the
density and temperature ratios between the centre of the star and the outer edge of the
radiative zone ($r = 0.55R$) are
$\overline{\rho}_\mathrm{centre}/\overline{\rho}_\mathrm{edge} \approx
4.15$, and $\overline{T}_\mathrm{centre}/\overline{T}_\mathrm{edge} \approx 1.6$.


\subsection{Dynamo solutions \label{dynamo-solutions}}

All of our runs, except for MHDr1, host a dynamo. In the corresponding
Run~zMHDr1 the magnetic field grows but the dynamo is very close
  to marginal, and
it is impractical to run it to saturation.

\subsubsection{Core dynamos \label{core-dynamos}}

\begin{figure*}[h!]
    \centering
    \includegraphics[scale=0.42]{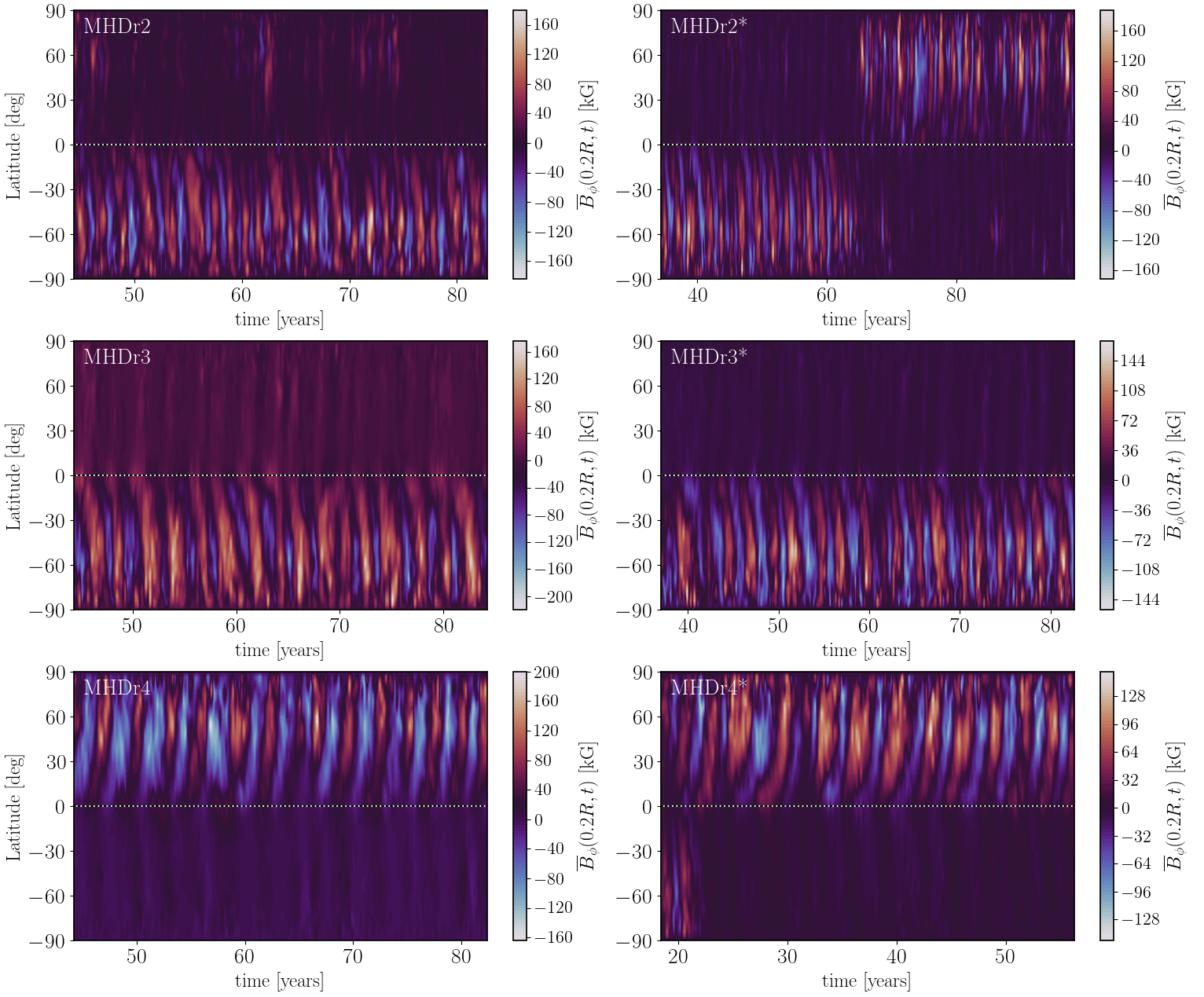}
    \caption{Time-latitude diagrams of the azimuthally averaged
      toroidal magnetic field $\overline{B}_\phi (r=0.2R,\theta,t)$ of
      the full star models. The run is indicated in the upper
      left corner of each panel.}
    \label{fig:B}
\end{figure*}

\begin{figure*}[h!]
    \centering
    \includegraphics[scale=0.42]{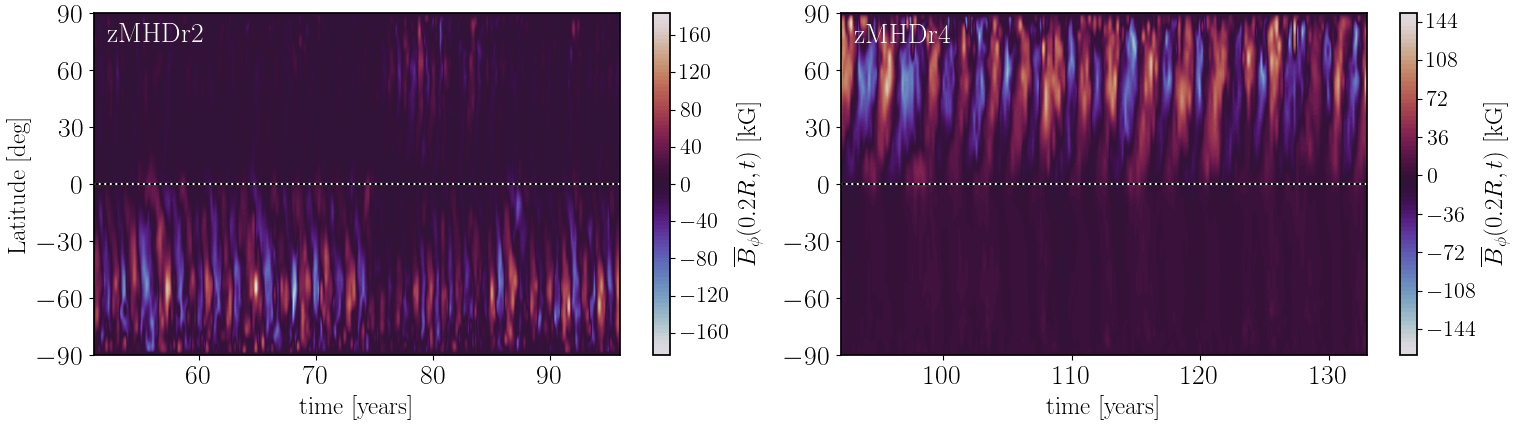}
    \caption{Same as Figure~\ref{fig:B} but for the zoom models zMHDr2
      and zMHDr4.}
    \label{fig:Bz}
\end{figure*}


\begin{figure*}[t!]
    \centering
    \includegraphics[width=\hsize]{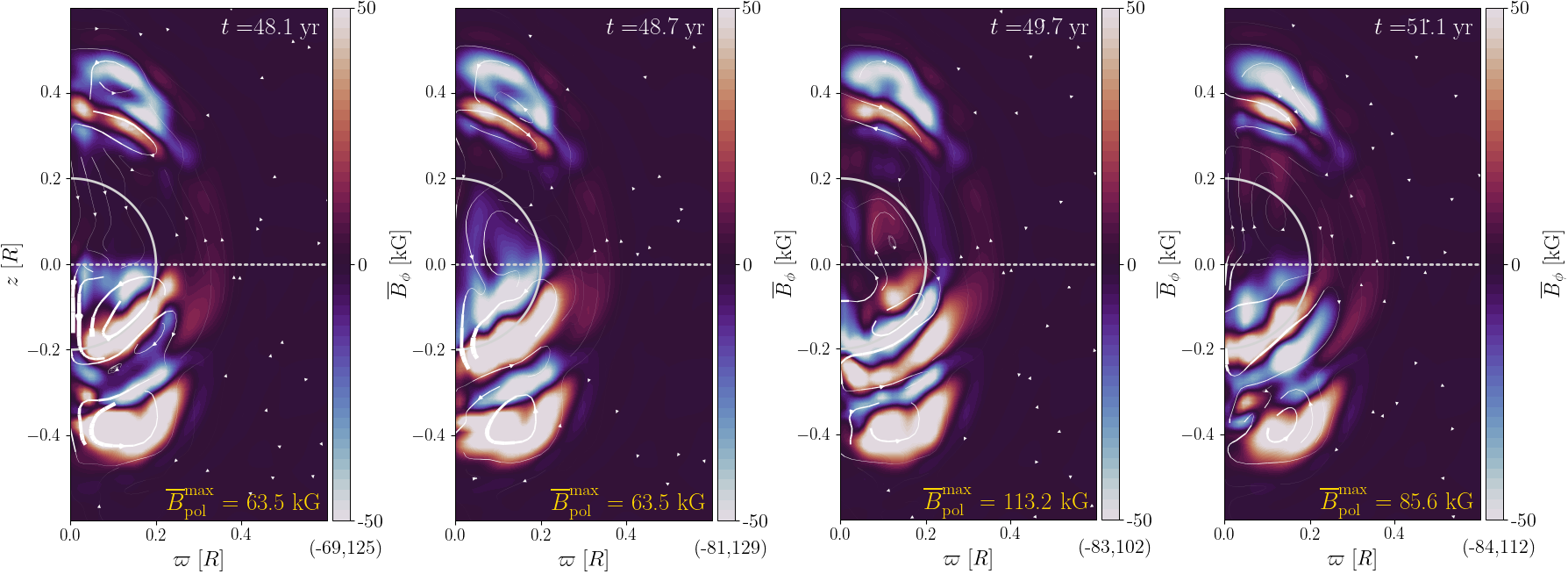}
    \caption{Azimuthally averaged toroidal magnetic field $\overline{B}_\phi(\varpi,z)$ from one cycle ($\sim$ 3 years) of MHDr3*. The poloidal magnetic field is represented with arrows, where the width is proportional to the strength of the field. The values of $\overline{B}_\phi$ are clipped to $\pm 50$ kG, and the maximum and minimum values $(\overline{B}_\phi^\mathrm{min},\overline{B}_\phi^\mathrm{max})$ are indicated below the colorbar. The dashed line at $z=0$ represents the equator.}
    \label{fig:Bpol}
\end{figure*}

The azimuthally averaged toroidal magnetic fields at $r=0.2R$ as
functions of time and latitude of the full star simulations are shown in
Figure~\ref{fig:B}. All of the magnetic field solutions are cyclic and
hemispheric. These are the first cyclic solutions reported from core
dynamos of early-type stars. Cyclic solutions have been
reported in simulations of fully convective stars where the geometry
of the dynamo region is similar to that of the current models with
similar
Coriolis number, $\mathrm{Co \approx 9}$ \citep{Kapyla-2021,
  ortizrodriguez-2023}. The main difference is the hemispheric nature
of the current simulations, although during limited periods a run in
\cite{Kapyla-2021} shows a predominantly hemispheric magnetic
field as well (see their Fig.~10). Here,
all the simulations show hemispheric fields most of the running time,
with equally strong fields on both hemispheres only rarely (in
particular Runs~MHDr2, MHDr4* and zMHDr2). Hemispheric dynamo
  solutions have been also found in Boussinesq
  \cite[][]{2011PEPI..185...61L,2012PhyS...86a8407S} and anelastic spherical shells models
  \cite[][]{2014A&A...567A.107R, 2016JFM...799R...6R}. In Runs~MHDr2
and MHDr3
the magnetic field is concentrated in the southern
hemisphere, while in MHDr4 the magnetic activity is located in the
northern hemisphere. In MHDr2*, the magnetic field starts in the
southern hemisphere, same as in MHDr2, but then it moves to the
northern hemisphere. This is a unique behaviour in this run and it is
unclear why it happens. Similar behaviour was also found by
\cite{brown-2020} in spherical simulations of fully convective M
dwarfs, with no conclusive explanation either.
The
rest of the MHD* runs are very similar to their MHD counterparts, as
shown in
Figure~\ref{fig:B}. The dynamo of MHDr3* is essentially the same as
that of MHDr3. MHDr4* has a cyclic dynamo on both hemispheres in the
early part of the simulation, but the activity in the southern
hemisphere vanishes after $\sim 23$ years, and the resulting
magnetic field is similar to MHDr4. Something similar was found by
\cite{brown-2020}, where their hemispheric dynamos are initially
symmetric on both hemispheres, but after some time strong
southern-hemisphere fields replace the original symmetric
configuration. The rms-magnetic fields from group MHD* are somewhat
weaker than in group MHD; see Table~\ref{table-1}.

Figure~\ref{fig:Bz} shows $\overline{B}_\phi(r=0.2R,\theta,t)$ for two
of the zMHD runs. These dynamos are also
very similar to their MHD group counterparts. zMHDr2 shows a dynamo in
the southern hemisphere. However, roughly between 75 and 85 years, it
exhibits some activity in the northern hemisphere as well, which
slightly reduces the activity in the southern hemisphere. This
behaviour is also somewhat more frequently visible in MHDr2, and it is
possibly related with the switch of active
hemisphere seen in
MHDr2*. The activity is almost completely hemispherical in all of the
other simulations in the zMHD group. The toroidal magnetic field of
zMHDr3 is not shown because
it is almost identical to its full star counterparts, although, the
mean intensity is somewhat weaker. Fourth column of
Table~\ref{table-1} shows
that a similar trend is present in all of the zMHD runs, so that
$B_\mathrm{rms}$ is about $10$ kG weaker than in the full star MHD
counterparts. As mentioned, the dynamo from zMHDr4 is very similar to
that of MHDr4. However, their magnetic cycles does not seem to match
and $P_\mathrm{cyc}$ from zMHDr4 is significantly shorter (2.50~yr)
than that of MHDr4 (3.14~yr). The strengths of the toroidal
  and radial magnetic fields at the surface of the convective core
  ($r=0.2R$) for the full star runs are
  $\overline{B}_\phi^\mathrm{rms} = 30-37$ kG and
  $\overline{B}_r^\mathrm{rms} = 16-19$ kG, while the zoom have
  $\overline{B}_\phi^\mathrm{rms} = 18-24$ kG and
  $\overline{B}_r^\mathrm{rms} = 9-13$ kG, respectively.
\\
\begin{figure*}[t!]
    \centering
    \includegraphics[width=\hsize]{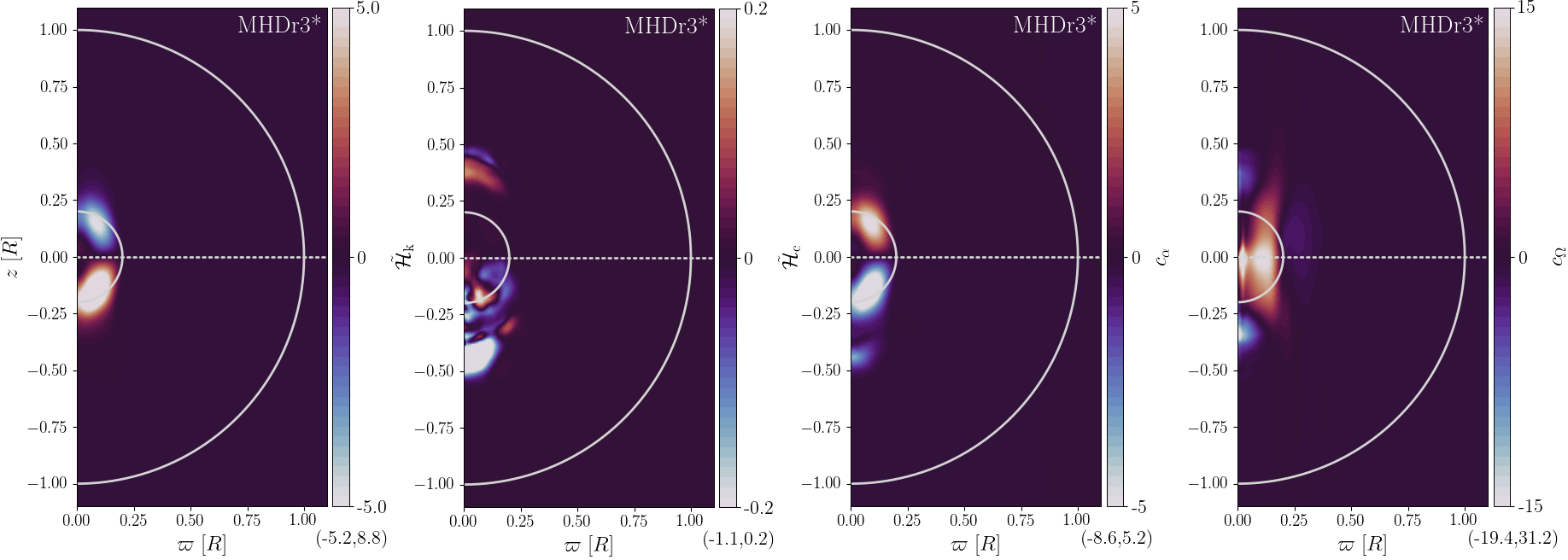}
    \caption{Time-averaged helicities and dynamo parameters from
      MHDr3*. From left to right the panels correspond to the
      normalised fluctuating kinetic helicity $\tilde{\mathcal{H}}_\mathrm{k} =
      \mathcal{H}_\mathrm{k}/(u_{\rm rms}^2/\Delta r)$, current helicity
      $\tilde{\mathcal{H}}_\mathrm{c} = \mathcal{H}_\mathrm{c}/(\rho_0 u_{\rm
        rms}^2/\Delta r)$ helicities, and the dynamo parameters
      $c_\alpha = \alpha \Delta r/\eta_\mathrm{t}$ and $c_\Omega=\partial \overline{\Omega}/\partial r (\Delta r)^3/\eta_\mathrm{t}$. All the panels are clipped for better
      legibility.}
    \label{fig:H&param}
\end{figure*}

The meridional distribution of the azimuthally averaged 
magnetic field of MHDr3* is shown in Figure~\ref{fig:Bpol}.
The amplitude of the mean toroidal component is typically 
larger than that of the mean poloidal component. 
  In the core the full star runs
  have rms values of $\overline{B}_{\rm rms}^{\rm tor}$
  = 20-25~kG and $\overline{B}_{\rm rms}^{\rm pol}$ = 18-22~kG, 
  while the zoom models have $\overline{B}_{\rm rms}^{\rm tor}$ = 
  17-20~kG and $\overline{B}_{\rm rms}^{\rm pol}$ = 15-17~kG. Although
  the magnetic activity in the core is
  hemispheric,
  there is still some magnetic field in the radiative zone above the
  non-active hemisphere of the core. From Fig.~\ref{fig:Bpol}, it is
  visible that this magnetic field might be coming from the core
  dynamo. At the beginning of the cycle ($t=48.1$ yr), the northern
  hemisphere of the core is almost free of magnetic activity. Briefly
  after, at $t= 48.7$ yr, some of the magnetism from the core dynamo
  migrates through the northern hemisphere. This
  magnetism reaches the lower part of the radiative envelope
  while the core dynamo changes its polarity. Once the cycle finishes,
  the northern hemisphere of the core gets almost devoid of magnetic
  activity again. The zone where the magnetic field is concentrated
  in the radiative zone
  starts around $r\approx 0.6R$, which is roughly the radius
  where the radial jumps of the diffusivities occur in the MHD*
  group. Therefore, due to the low diffusivities in these layers, the
  magnetic field can stay and evolve in long timescales compared to
  the period of the core dynamo. These magnetic fields might be
transported from the core to the bottom of the radiative
  envelope by the columnar flows produced
by the Taylor-Proudman theorem, due to the high Coriolis numbers in
our simulations (see Section~\ref{flows}). The reason of why these
vertical flows can penetrate the stable stratification of the
radiative envelope could be the unrealistically low Richardson number
computed in such layers (see Table~\ref{table-1}). These low values
are a direct consequence of the low Brunt-Väisälä frequencies achieved
by our simulations, as in real stars these values are expected to be
several of orders of magnitudes higher. In the southern hemisphere 
the situation is similar, but the magnetic field has a different 
polarity than in the north.

To understand the origin of the dynamos in the simulations we
  make estimates of commonly used dynamo numbers in mean-field dynamo
  theory \citep[e.g.][]{KR80}. This involves computing the $\alpha$
  effect which is proportional to the kinetic helicity of the flow
  \citep[][]{1966ZNatA..21..369S}.
The fluctuating kinetic and current helicities are
\begin{align}
    \mathcal{H}_\mathrm{k} &= \overline{\bm{U} \bm{\cdot} \bm{\omega}} - \overline{\bm{U}} \bm{\cdot} \overline{\bm{\omega}}, &
    \mathcal{H}_\mathrm{c} &= \overline{\bm{J} \bm{\cdot} \bm{B}} - \overline{\bm{J}} \bm{\cdot} \overline{\bm{B}},
\end{align}
respectively, where $\bm{\omega} = \bm{\nabla} \times \bm{U}$ is 
the vorticity. Furthermore, to quantify the $\alpha$ and
the $\Omega$ effects, it is convenient to introduce the following
dynamo parameters \citep{Kapyla-2013}
\begin{align}
    c_\alpha = \frac{\alpha \Delta r}{\eta_\mathrm{t}},\hspace*{0.5cm} c_\Omega = \frac{\partial \overline{\Omega}/\partial r (\Delta r)^3}{\eta_\mathrm{t}}, \label{dyn-param}
\end{align}
where $\eta_\mathrm{t} = \tau u^2_\mathrm{rms}/3$ is an estimate of
the turbulent
diffusivity and $\tau = \Delta r/u_\mathrm{rms}$ is the convective
turnover time. Furthermore, the non-linear $\alpha$ effect is given by
\citep{PFL76}
\begin{equation}
\alpha = -\frac{\tau}{3}\left({\cal H}_k - {\cal H}_c/\overline{\rho} \right).
\end{equation}
Finally, $\overline{\Omega}$ is the time- and azimuthally-averaged
rotation rate.

The normalized fluctuating kinetic helicity
$\tilde{\mathcal{H}}_\mathrm{k} = \mathcal{H}_\mathrm{k}/(u_{\rm
rms}^2/\Delta r)$ and current helicity  $\tilde{\mathcal{H}}_\mathrm{c} =
\mathcal{H}_\mathrm{c}/(\rho_0 u_{\rm rms}^2/\Delta r)$
of MHDr3*, as well as the dynamo parameters (\ref{dyn-param}) are shown
in Figure~\ref{fig:Bpol}. The kinetic helicity is negative (positive)
in the northern (southern) hemisphere. The current helicity has a
less well defined large-scale structure and a much lower amplitude.
Along with the solar-like differential rotation of the core, which is
discussed in Section~\ref{flows}, poleward propagation of dynamo waves
is expected in the $\alpha\Omega$ dynamo approximation
\citep[e.g.][]{Pa55b,1975ApJ...201..740Y}. This is consistent with
Figures~\ref{fig:B} and \ref{fig:Bpol}. Furthermore, as shown in the
rightmost panels of
Figure~\ref{fig:H&param}, the values of $c_\Omega$ inside the core, are
larger than those of $c_\alpha$. Although the difference between
$\overline{B}_{\rm rms}^{\rm tor}$ and $\overline{B}_{\rm rms}^{\rm
  pol}$ is not very large, we nevertheless interpret the dynamos in
our simulations to be of some $\alpha\Omega$ or $\alpha^2\Omega$
flavour. We note that $c_\alpha$ is close to zero at the base of the
radiative zone between $z=0.25R$ and $z=0.5R$, which suggests lack of
dynamo action there.
In the southern hemisphere between $z=-0.5R$ and $z=-0.25R$,
$c_\alpha$ has non-zero values, but these
do not overlap significantly with the non-zero values of
$c_\Omega$, suggesting dynamo action unlikely as well.
Therefore, the interpretation that
the magnetic field in the radiation zone, as seen in
Figure~\ref{fig:Bpol}, is transported there from the core dynamo
remains plausible. 
The rest of the simulations have very similar kinetic helicity and
dynamo coefficient profiles regardless of the location of the core
dynamo.

Magnetic helicity conservation has profound consequences for
  large-scale dynamos. If magnetic helicity cannot exit the system,
  the $\alpha$ effect can be catastrophically quenched leading to
  resistively slow large-scale magnetic field growth
  \citep[e.g.][]{B01}. In solar-like stars the dynamo-active region
  can shed magnetic helicity to the surrounding interstellar space via
  coronal mass ejections and other eruptive events. No such
  possibility exists for massive stars where the core dynamo is
  isolated from the stellar surface. A plausible possibility in this
  case is a diffusive magnetic helicity flux toward the equator where
  oppositely signed helicity can cancel. Such scenario has been
  demonstrated in simpler forced turbulence simulations where the
  kinetic helicity changes sign at the equator similarly as in the
  current models \citep[e.g.][]{2010AN....331..130M}. However, we
  postpone further investigation of this issue to future studies.

\subsubsection{Surface magnetic fields \label{surfB}}

\begin{figure}[t!]
    \centering
    \includegraphics[width=\hsize]{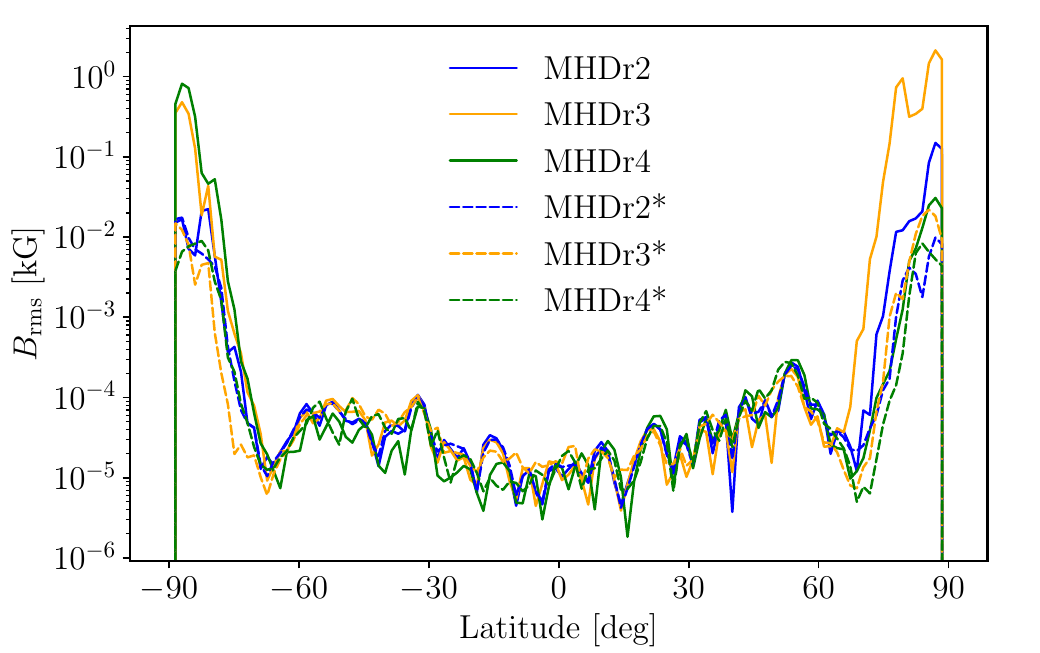}
    \caption{Time averaged rms magnetic field at the surface of the
      star $r=R$ as a function of latitude from the runs in groups
      MHD and MHD*.}
    \label{fig:Bsurf}
\end{figure}

Figure~\ref{fig:Bsurf} shows the rms magnetic field $B_\mathrm{rms}$
at the surface of the star for the MHD and MHD* simulations.
As expected, the magnetic field generated by the core dynamo is unable
to create strong large-scale magnetic structures at the
surface of the star. In all
simulations the magnetic field is nearly zero on almost the whole
surface of the star, except at the poles. The $B_\mathrm{rms}$ from
latitudes between $-90$ and $-75$
degrees and from $75$ to $90$ degrees, that is near the poles,
have the order of
$0.1$~kG in the MHD group, and $10^{-3}$~kG in the MHD* group. While
the magnetic fields from the rest of latitudes in both groups have
values of the order of $10^{-5}$ kG. Additionally, there does not seem
to be any correlation between the amount of magnetic flux that reaches
the surface and the rotation period of the star. The highest rms
value averaged over the poles
in the MHD group comes from MHDr3 with $0.47~\mathrm{kG}$,
and in the MHD* group from MHDr3*, with $6.8 \cdot
10^{-3}~\mathrm{kG}$.
Similarly to what happens in the bottom radiative envelope,
  the magnetic field are probably transported from the core to the
  surface near the poles
  by the combination of unrealistic Richardson numbers and axially
  aligned flows due to the high rotation of the simulations.

Although the radial profiles of the diffusivities in the MHD* group
reduce the intensity of these flows and therefore, the
magnetic field in the poles, they are
still two orders of magnitude stronger than the magnetic fields in the
rest of the stellar surface. In principle, increasing the Brunt-Väisälä frequency closer to realistic values should reduce the spreading of flows from the core to the envelope \citep[see e.g.][]{Lydia-2024}. However, increasing this frequency is numerically very expensive, therefore
with the current resolution of our simulations these results at the
stellar surface should be taken with caution.

Finally, the magnetic field that reaches the poles is not only
  located at the pole corresponding to the active hemisphere despite
  the hemispheric nature of the core dynamo. This can be a
  consequence of the distribution seen in
  Fig.~\ref{fig:Bpol}, as the flows might be transporting a 
  fraction of the magnetic flux from the base of the bottom radiative
  envelope. The local rms velocity $u_\mathrm{rms}(\varpi,z)$
  averaged from $\varpi=0$ to $\varpi = 0.2R$, that is, close to the
  axis of rotation, has comparable values from $z=0.2R$ to $z=R$
  and from $z=-0.2R$ to $z=-R$. Therefore a fraction
  of the magnetic field in the radiative zone seen in 
  Fig.~\ref{fig:Bpol} is transported to the corresponding pole.
  From Fig.~\ref{fig:Bsurf} is visible that the pole
  over the inactive hemisphere has a $B_\mathrm{rms}$ typically an 
  order
  of magnitude weaker than that of the other pole, but it is still
  significantly higher than the rest of latitudes. The relatively
  high values of $u_\mathrm{rms}(\varpi,z)$ close to the axis of
  rotation are likely due to the combination of much higher diffusion
  coefficients and lower Richardson number in our simulations than in
  real stars. The advection timescales of the vertical flows are 
  estimated to be around 100~yrs in the MHD group and 300~yrs in
  the MHD* group, while diffusion timescales are around $20$~yrs in
  both groups. Therefore, the magnetic field is transported to
  the surface most likely due to turbulent diffusion.

\subsection{Large-scale flows \label{flows}}

The time- and azimuthally-averaged rotation rate is given by:
\begin{equation}
    \overline{\Omega}(\varpi, z) = \Omega_0 + \overline{U}_\phi (\varpi, z) / \varpi, \label{average-rot}
\end{equation}
where $\varpi = r \sin \theta$ is the cylindrical radius. Furthermore,
the averaged meridional flow is:
\begin{equation}
    \overline{\bm{U}}_\mathrm{mer}(\varpi,z) = (\overline{U}_\varpi,0,\overline{U}_z).  \label{average-mer-flow}
\end{equation}
We use the same parameters as in
\cite{Kapyla-2013}, \cite{Kapyla-2021}, and \cite{ortizrodriguez-2023}
to quantify the amplitude of the radial and latitudinal differential
rotation. These are given by:
\begin{align}
    \Delta_\Omega^{(r)} &= \frac{\overline{\Omega}(r_\mathrm{top}, \theta_\mathrm{eq}) - \overline{\Omega}(r_\mathrm{bot}, \theta_\mathrm{eq}) }{\overline{\Omega}(r_\mathrm{top}, \theta_\mathrm{eq})}, \label{rot-param1}\\ \Delta_\Omega^{(\overline{\theta})} &= \frac{\overline{\Omega}(r_\mathrm{top}, \theta_\mathrm{eq}) - \overline{\Omega}(r_\mathrm{top}, \overline{\theta}) }{\overline{\Omega}(r_\mathrm{top}, \theta_\mathrm{eq})}, \label{rot-param2}
\end{align}
where $r_\mathrm{top} = 0.9R$ and $r_\mathrm{bot} = 0.1R$ are the
radius near the surface and centre of the star,
respectively. $\theta_\mathrm{eq}$ corresponds to the latitude at the
equator, and $\overline{\theta}$ is an average of $\overline{\Omega}$
between latitudes $-\theta$ and $\theta$. Additionally, it is
relevant to also analyse the differential rotation of the convective
core. Therefore, we introduce $\Delta_\Omega^{\mathrm{CZ}(r)}$ and
$\Delta_\Omega^{\mathrm{CZ} (\overline{\theta})}$, which are
the same as (\ref{rot-param1}) and (\ref{rot-param2}), but with
$r_\mathrm{top}=0.2R$ and $r_\mathrm{bot}=0.05R$.

\begin{table*}[t!]
\centering
\caption{Differential rotation parameters and the maximum meridional flow $\overline{U}_\mathrm{mer}^\mathrm{max}$ from all simulations.
}
\begin{tabular}{lccccccc}
\hline\hline\noalign{\smallskip}
Run & $\Delta_\Omega^{(r)}$ & $\Delta_\Omega^{(\overline{\theta})}(60^\circ)$ & $\Delta_\Omega^{(\overline{\theta})}(75^\circ)$ & $\Delta_\Omega^{\mathrm{CZ}(r)}$ & $\Delta_\Omega^{\mathrm{CZ}(\overline{\theta})}(60^\circ)$ & $\Delta_\Omega^{\mathrm{CZ}(\overline{\theta})}(75^\circ)$ & $\overline{U}_\mathrm{mer}^\mathrm{max}~\mathrm{[m/s]}$ \\
\hline\noalign{\smallskip}
MHDr1   & 0.1579 &  -0.0001 &  -0.0011 &  0.3801  &  0.2650  &  0.4073 & 7.7 \\
MHDr2   &  0.0505  & -0.0001  & -0.0006 &  0.1131  & 0.0699  &  0.1142 & 6.0\\
MHDr2*I  & 0.0405 &  -0.0001  & -0.0007  &   0.0941 &   0.0607    & 0.0937 & 3.2\\
MHDr2*II  & 0.0538  & -0.0000  & -0.0006  &   0.1209  &  0.0734   & 0.1265 & 3.3\\
MHDr3  & 0.0251  & -0.0000  & -0.0004 &  0.0538 & 0.0295  & 0.0484 & 3.2 \\
MHDr3*   & 0.0224  & -0.0001  & -0.0006  &   0.0489  & 0.0283  &  0.0438 & 2.1 \\
MHDr4   & 0.0197  & -0.0000 & -0.0003 &  0.0405 & 0.0213  & 0.0350 & 2.5 \\
MHDr4*  & 0.0169  & -0.0001  & -0.0006 &  0.0387 & 0.0208  & 0.0333 & 2.3\\
\hline
zMHDr1  & - & - & - & 0.3742  &  0.2442  &  0.3729 & 7.8\\
zMHDr2  & - & - & - & 0.1290 &  0.0739  &  0.1294 & 6.4 \\
zMHDr3 & - & - & - & 0.0511 &  0.0282  & 0.0440 & 4.4\\
zMHDr4 & - & - & - & 0.0434  &  0.0214  &  0.0377 & 2.7  \\
\hline

\end{tabular}
\label{table-DR2}
\tablefoot{Run MHDr2* includes the parameters before (I) and after (II) the dynamo migration. The differential rotation parameters follow Eqs.~(\ref{rot-param1}) and (\ref{rot-param2}).}
\end{table*}

\begin{figure}[t!]
    \centering
    \includegraphics[scale=0.55]{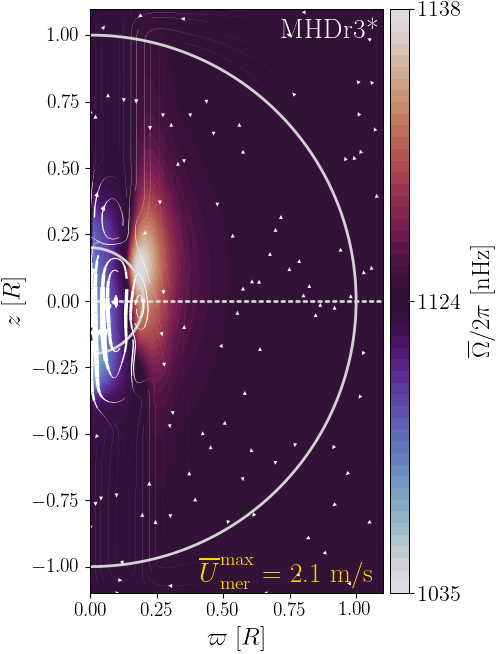}
    \caption{Profile of the temporally and azimuthally averaged
      rotation rate $\overline{\Omega}(\varpi,z)$ for MHDr3*. The
      streamlines indicate the mass flux due to meridional circulation
      and the maximum averaged meridional flow is indicated in the
      lower right side of the plot. The dashed line represents the
      equator.}
    \label{fig:omegafull}
\end{figure}

The differential rotation parameters are summarised in
Table~\ref{table-DR2}. From columns
two to four of Table~\ref{table-DR2}, we can conclude that the stellar
surface rotates nearly rigidly, because
$|\Delta_\Omega^{\overline{\theta}}|$ is $\sim 10^{-4}$ for 
$\theta=60^\circ$ and $\theta=70^\circ$.
This implies negligible difference between the averaged rotation at
the chosen latitudes and the equator. Figure~\ref{fig:omegafull} shows
that in run MHDr3*, rigid rotation extends from the surface to a
significant part of the radiative envelope between $r\approx 0.5R$ and
$r=R$.
This is similar
to the radiative interior of the Sun, which has approximately rigid
rotation (see \citealt{Rachel-2009}). This behaviour is present in all 
the simulations from this study. Unlike the simulations by
\cite{Augustson-2016} whose radiative envelopes are almost completely
rigid (see their Fig.~7), the current runs show a stronger
differential rotation in the regions closest to the core.

\begin{figure*}[t!]
    \centering
    \includegraphics[scale=0.43]{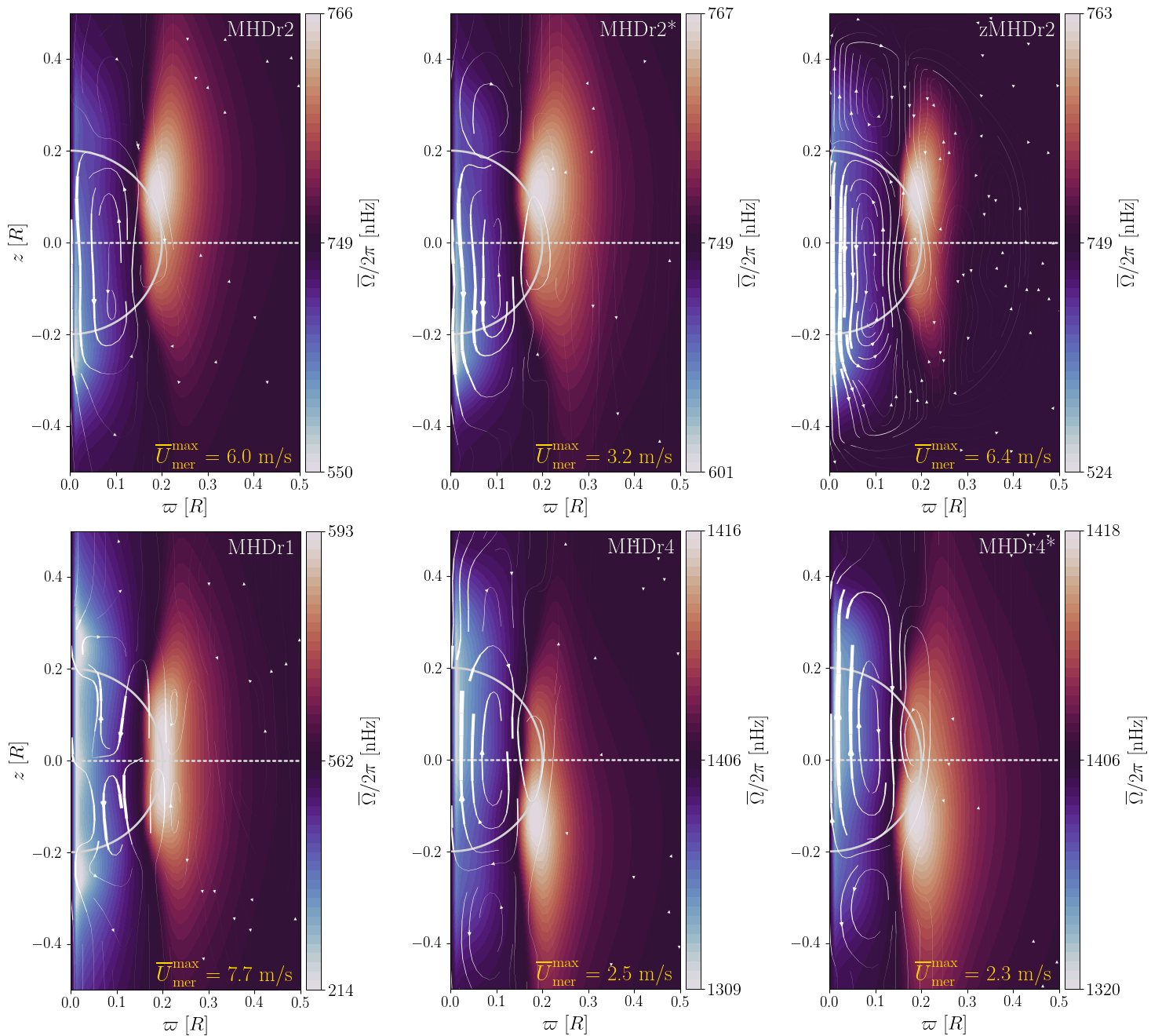} 
    \caption{Profiles of the temporally and azimuthally averaged
      rotation rate $\overline{\Omega}(\varpi,z)$ of selected runs
      from all sets (clipped at $r = 0.5R$). The upper panels show all
      the runs with $P_\mathrm{rot}= 15~\mathrm{days}$ (MHDr2, MHDr2*
      and zMHDr2). The lower panels shows the run with no dynamo
      (MHDr1), and two simulations with
      $P_\mathrm{rot}=8~\mathrm{days}$ (MHDr4 and MHDr4*).}
    \label{fig:omega}
\end{figure*}

All runs have positive $\Delta_\Omega^{(r)}$ because the
innermost part of the core is rotating slower than the stellar
surface, as shown in Figure~\ref{fig:omegafull}. In
Figure~\ref{fig:omega} the rotation profiles of other representative
simulations are shown. The radial extent is now limited to
$0.5R$, allowing a better display of the flows from the core and
the portion of the radiative envelope with non-rigid rotation. The
convective cores from all the runs have solar-like differential
rotation, similar to the solar-like state reported by
\cite{brown-2020}, and the hydrodynamic case H4 of 
\cite{Augustson-2016}. This is also shown by the positive values of
$\Delta_\Omega^{\mathrm{CZ}(\overline{\theta})}$ in
Table~\ref{table-DR2}, which indicate that the convective core is
rotating faster at the equator than at other latitudes. Furthermore,
we can see vertical structures that are parallel to the axis of
rotation due to the Taylor-Proudman theorem. This is similar to
the fast rotator from \citealt{Brun-2005} (see their
Fig.~9). As mentioned in Section~\ref{surfB}, these flows penetrate
the radiative zone probably as a consequence of the low
Brunt-V\"ais\"al\"a
frequencies in our simulations. We estimate that ${\rm
    Ri}_\Omega$ in a real A0 star is of the order of $10^4$ which is
  five to six orders of magnitude higher than in the current
  simulations (see the tenth column in Table~\ref{table-1}). 
  Higher values of $N$ would also
increase the ratio between the Eddington-Sweet and viscous timescales
\citep[e.g.][]{2012ApJ...755...99W}, and the mean flows would only
penetrate a small distance outside the core \citep[see, e.g. Fig.~5
  of][]{Lydia-2024}. 
  Therefore, although high rotation rates influence the
  efficiency of angular momentum transport in massive stars 
  \citep{2019ARA&A..57...35A},
  we would not expect differential rotation to spread in such a
  significant part of the radiative envelope if realistic stellar
  parameters were used.

The Run~MHDr1 without a dynamo has the largest differential rotation
in the core; see columns five to seven in Table~\ref{table-DR2}. The
rotation profile is symmetric with respect to the equator; see the
left lower panel of Figure~\ref{fig:omega}. On the other hand, MHDr2
and MHDr4 are asymmetric with respect
to the equator. This is due to the presence of strong hemispheric
dynamos, which quench the differential rotation where the magnetic
fields are strong. Magnetic quenching of differential rotation has 
been reported by various other numerical simulations 
\citep[e.g.][]{Brun2004, Kapyla2017, Bice2023}.
Run~MHDr3 behaves similarly to MHDr2. MHD* group shows similar
  behaviour.
The rotation profile of MHDr2* shown in Figure~\ref{fig:omega} is
averaged between 13 and 60 years, that is, before the dynamo moves from
the southern to the northern hemisphere. As a consequence, the
rotation profile is asymmetric with respect to the equator and very
similar to that of MHDr2. After the dynamo migration ($t > 60$ yrs)
the rotation profile is inverted with respect to the equator and
resembles that of MHDr4.
The rest of the runs in the MHD* group are
almost identical to their MHD counterparts, which is shown for
Runs~MHDr4* and MHDr4 in the lower panels of
Figure~\ref{fig:omega}. We note that the maximum meridional velocities
$\overline{U}_\mathrm{mer}^\mathrm{max}$ are always smaller in the MHD*
group (see last column of Table~\ref{table-DR2}). Here the
  meridional flow amplitude is not monotonic in the rotation rate as
  the value of MHDr4* is slightly higher than that of MHDr3*, whereas
in the MHD group the meridional
flow amplitude decreases with increasing rotation rate.

The rotation profiles from group zMHD are very similar to their MHD
counterparts, with similar maximum and minimum values of
$\overline{\Omega}/2\pi$, and slightly higher maximum values of the
averaged meridional flow, as shown for zMHDr2 in the upper right
panel of Figure~\ref{fig:omega}. zMHDr1 has
$\overline{U}_\mathrm{mer}^\mathrm{max} = 7.8~\mathrm{m/s}$ being the
highest value of the simulations. 
In the rotation profile of zMHDr2, we
can clearly distinguish more structures than in that of MHDr2. This is
due to the increase of grid points inside the convective core,
allowing to resolve flows on smaller scales. The same happens with all
the runs from the zMHD group, but the main results remain the same,
and therefore they are not included in the plot. In this set of
simulations $\overline{U}_\mathrm{mer}^\mathrm{max}$ decreases with
increasing $\Omega_0$, similar to the MHD group.

\subsection{Energy analysis}
\subsubsection{Global and core energies}
The total kinetic and magnetic energies, respectively, are given by
\begin{align}
    E_\mathrm{kin} &= \frac{1}{2} \int  \rho \bm{U}^2  dV, & E_\mathrm{mag} &= \frac{1}{2\mu_0} \int  \bm{B}^2  dV.
    \label{energies-1}
\end{align}
The
energies for the differential rotation (DR) and meridional circulation
(MC), are
\begin{align}
    E_\mathrm{kin}^\mathrm{DR} &= \frac{1}{2} \int \rho \overline{U}_\phi^2  dV, & E_\mathrm{kin}^\mathrm{MC} &= \frac{1}{2} \int  \rho (\overline{U}_\varpi^2 + \overline{U}_z^2)  dV. \label{energies-2}
\end{align}
Furthermore, the toroidal and poloidal magnetic energies are defined
as
\begin{align}
    E_\mathrm{mag}^\mathrm{tor} &= \frac{1}{2\mu_0} \int  \overline{B}_\phi^2 dV, & E_\mathrm{mag}^\mathrm{pol} &= \int \frac{1}{2\mu_0} (\overline{B}_\varpi^2 + \overline{B}_z^2)  dV. \label{energies-3}
\end{align}
These expressions (\ref{energies-1}-\ref{energies-3}) were integrated 
over the volume of the star ($r < R$), or over the convective 
core ($r < \Delta r$), to study the energies in the entire 
star and in core, respectively. The obtained values are listed in 
Table~\ref{table-energy}.

\begin{table*}[h!]
\centering
\caption{Kinetic and magnetic energies.}
\begin{tabular}{lccccccc}
\hline\hline\noalign{\smallskip}
Run & $E_\mathrm{kin}~[10^{33}\mathrm{J}]$  & $E_\mathrm{kin}^\mathrm{DR}/E_\mathrm{kin}$ & $E_\mathrm{kin}^\mathrm{MC}/E_\mathrm{kin}$ & $E_\mathrm{mag}~[10^{33}\mathrm{J}]$  & $E_\mathrm{mag}/E_\mathrm{kin}$ & $E_\mathrm{mag}^\mathrm{tor}/E_\mathrm{mag}$ & $E_\mathrm{mag}^\mathrm{pol}/E_\mathrm{mag}$\\
\hline\noalign{\smallskip}
Full star \\
\hline
MHDr1  & 296.78  & 0.631  & 0.017  & -  & - & -  & - \\
MHDr2  & 74.50  & 0.398  & 0.017  & 24.00  & 0.322  & 0.294  & 0.061 \\
MHDr2*  & 74.70  & 0.391  & 0.018  & 24.71  & 0.331  & 0.277  & 0.059 \\
MHDr3  & 42.57  & 0.319  & 0.011  & 23.99  & 0.563  & 0.229  & 0.079 \\
MHDr3*  & 43.81  & 0.315  & 0.010  & 30.52  & 0.697  & 0.264  & 0.074 \\
MHDr4  & 34.86  & 0.323  & 0.008  & 23.79  & 0.682  & 0.300  & 0.086 \\
MHDr4*  & 35.84  & 0.317  & 0.006  & 26.58  & 0.742  & 0.266  & 0.069 \\
\hline
Core \\
\hline
MHDr1  & 32.64  & 0.576  & 0.025  & -  & -  & -  & - \\
MHDr2  & 8.88  & 0.346  & 0.026  & 1.74  & 0.196  & 0.140  & 0.097 \\
MHDr2*  & 8.95  & 0.341  & 0.027  & 1.56  & 0.174  & 0.132  & 0.098 \\
MHDr3  & 5.10  & 0.286  & 0.016  & 2.06  & 0.403  & 0.151  & 0.112 \\
MHDr3*  & 4.91  & 0.247  & 0.016  & 1.86  & 0.379  & 0.139  & 0.102 \\
MHDr4  & 4.19  & 0.309  & 0.011  & 1.74  & 0.415  & 0.189  & 0.118 \\
MHDr4*  & 3.96  & 0.259  & 0.009  & 1.58  & 0.399  & 0.172  & 0.108 \\
zMHDr1  & 27.47  & 0.537  & 0.032  & -  & -  & -  & - \\
zMHDr2  & 9.00  & 0.371  & 0.035  & 1.20  & 0.133  & 0.133  & 0.091 \\
zMHDr3  & 4.61  & 0.275  & 0.021  & 1.40  & 0.304  & 0.152  & 0.102 \\
zMHDr4  & 3.83  & 0.331  & 0.014  & 1.06  & 0.277  & 0.186  & 0.104 \\
\hline
\end{tabular}
\label{table-energy}
\tablefoot{The energies are averaged over time in the full
  star and in the convective core. The total kinetic and
  magnetic energies are given in units of $10^{33}\mathrm{J}$.}
\end{table*}

The slowest rotators have the highest kinetic energies and $\Ekin$
decreases with $\Omega_0$. This is a consequence of
high rotation rates reducing the intensity of the flows, as visible
from $u_\mathrm{rms}$ in Table~\ref{table-1} and the differential
rotation parameters in Table~\ref{table-DR2}. We note that MHDr1 has a
significantly higher value of $E_\mathrm{kin}$ than the rest of the
simulations, which is a consequence of the absence of dynamo in
this run. Earlier simulations
\citep[e.g.][]{2008ApJ...689.1354B,viviani-2018} and theoretical
considerations \citep[e.g.][]{KR99} suggest that differential rotation
and meridional circulation decrease when the rotation rate
increases. Our results are mostly in agreement with these
results, although the most rapidly rotating runs in each group deviate
from this trend. A possible explanation is that the flows penetrate
deeper in the radiative layer in these simulations due to the higher
rotation rate and smaller Richardson numbers.

\begin{figure}[t!]
    \centering
    \includegraphics[width=\hsize]{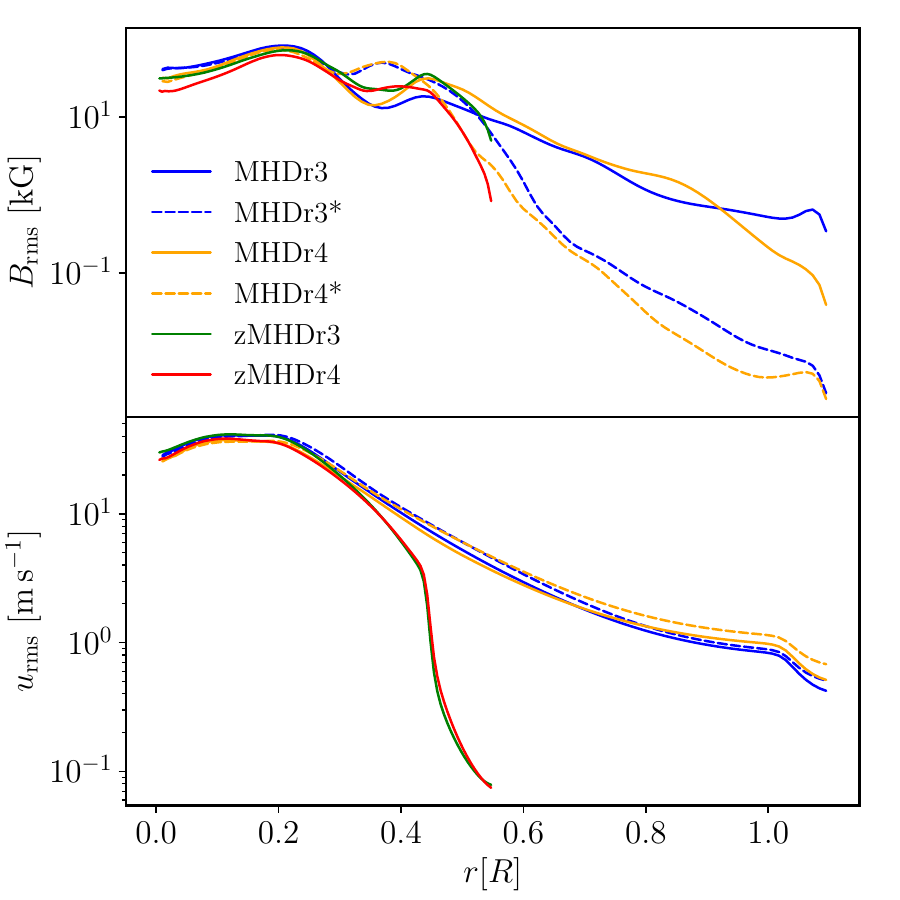}
    \caption{Radial profiles of the horizontally averaged
      $B_\mathrm{rms}$ (\textit{top panel}) and $u_\mathrm{rms}$
      (\textit{bottom panel}) from representative runs.}
    \label{fig:radBrms}
\end{figure}

The magnetic and kinetic energies of the entire star are
always larger in the group MHD* than in group MHD.
This is because of the different diffusivity profiles between
groups. In Figure~\ref{fig:radBrms} is visible from the $B_\mathrm{rms}$
profile that runs from group MHD* have a secondary maxima closer to
the core than
those of group MHD. These maxima above the core envelope interface
around $0.36R$ lead to higher overall values of $B_\mathrm{rms}$ in
the MHD* group even though the fields in the radiative envelope are
weaker in these cases.
However, this
behavior disappears in the core, as shown in the last rows of
Table~\ref{table-energy}. With $r< \Delta
r$ the values of the magnetic energy in group MHD* are smaller than in
group MHD, while the values of the kinetic energy, although quite
close, do not seem to show any tendency between groups. Similarly to
the differential rotation energies, the toroidal magnetic energies are
very similar, with values of
$E_\mathrm{mag}^\mathrm{tor}/E_\mathrm{mag}$ between 0.229 and 0.3 for
$r<R$, and between 0.132 and 0.189 for $r< \Delta r$. In both cases,
the highest value occurs in MHDr4, same as with the poloidal
magnetic energy. However, the maximum $E_\mathrm{mag}$ occurs in
MHDr3* for the entire star, and in MHDr3 for the core. This is
consistent with Table~\ref{table-1}, as these runs show the highest
$B_\mathrm{rms}$. 
In the core, toroidal magnetic energies are somewhat larger than
the poloidal energies, with typically $E_\mathrm{mag}^{\rm tor} \approx 1.5
E_\mathrm{mag}^{\rm pol}$. This is consistent with the discussion
about the nature of the core dynamo; see Section~\ref{core-dynamos}.

\begin{figure}[t!]
    \centering
    \includegraphics[width=\hsize]{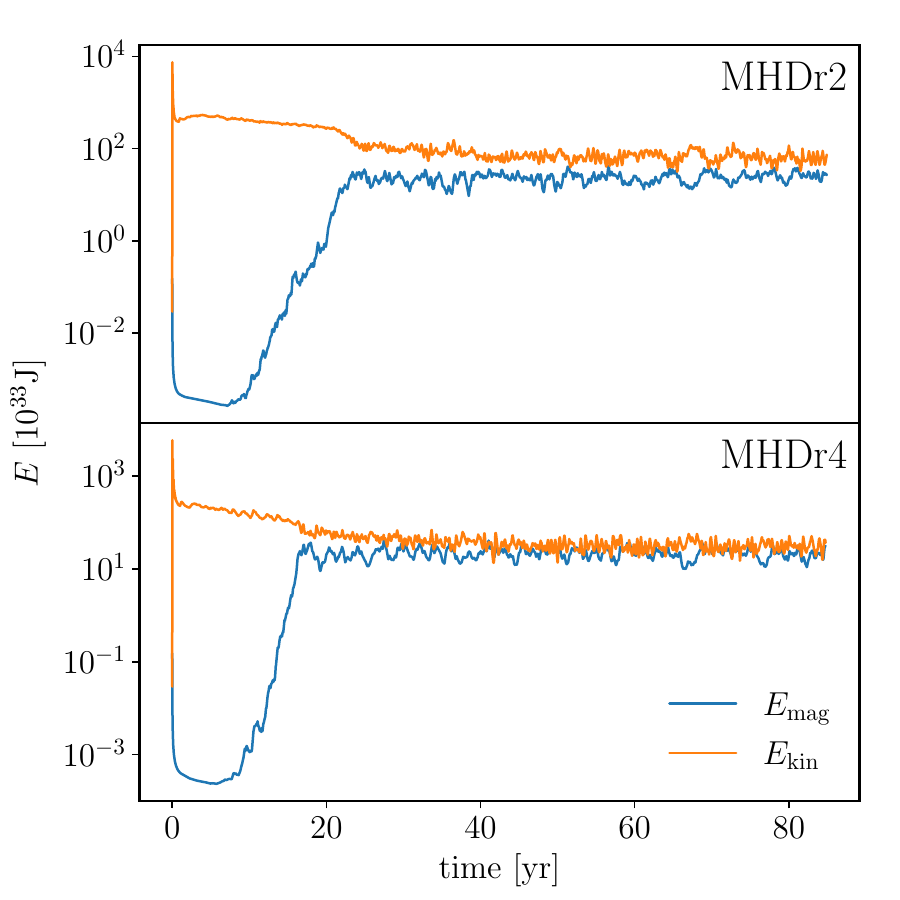}
    \caption{Temporal evolution of the magnetic (blue) and kinetic
      (orange) energies in the full star of the runs
      MHDr2 (\textit{top panel}) and MHDr4 (\textit{bottom panel}).}
    \label{fig:energies}
\end{figure}

The magnetic energies in all of the current simulations are below
  equipartition even in the most rapidly rotating cases, typically
  with $E_\mathrm{mag}/E_\mathrm{kin} \approx 0.4$, and only when
  considering the energies in the whole star the magnetic energy is
  near or at equipartition; see Fig.~\ref{fig:energies}. We therefore
  do not reach the near equipartition regime in the dynamo region that
  was found by \cite{Brun-2005}. This is likely because of the
  relatively laminar parameter regime explored in the current study.
The kinetic energies in the zMHD group are comparable to the
corresponding full star runs. However,
as with $B_\mathrm{rms}$, the total magnetic energy in the zMHD runs
are lower than in the rest of simulations.

\subsubsection{Energy fluxes and luminosities}
\label{sec:fluxes}

The luminosities related to the radiative, enthalpy, kinetic
energy, and viscous fluxes, and the additional cooling and heating,
are defined as \citep{Kapyla-2021}:
\begin{align}
    \mathscr{L}_\mathrm{rad} &= -4\pi r^2 \langle K \rangle_{\theta\phi t} \frac{\partial \langle T \rangle_{\theta\phi t}}{\partial r}, & \mathscr{L}_\mathrm{enth} &= 4\pi r^2 c_\mathrm{P} \langle(\rho U_r)' T' \rangle_{\theta\phi t}, \\
    \mathscr{L}_\mathrm{kin} &= 2\pi r^2 \langle \rho \bm{U}^2 U_r \rangle_{\theta\phi t}, & \mathscr{L}_\mathrm{visc} &= -8 \pi r^2 \nu \langle \rho U_i \mathsf{S}_{ir} \rangle_{\theta\phi t}, \\
    \mathscr{L}_\mathrm{cool} &= - \int_0^r 4\pi r^2 \langle \mathcal{C} \rangle_{\theta\phi t} dr, & \mathscr{L}_\mathrm{heat} &= \int_0^r 4\pi r^2 \langle \mathcal{H} \rangle_{\theta\phi t} dr,
\end{align}
where primes indicates fluctuations from the horizontal ($\theta\phi$)
average, $r$ denotes the radial component in spherical coordinates,
and $\langle . \rangle_{\theta \phi t}$
horizontal and temporal averaging.

\begin{figure}[t!]
    \centering
    \includegraphics[width=\hsize]{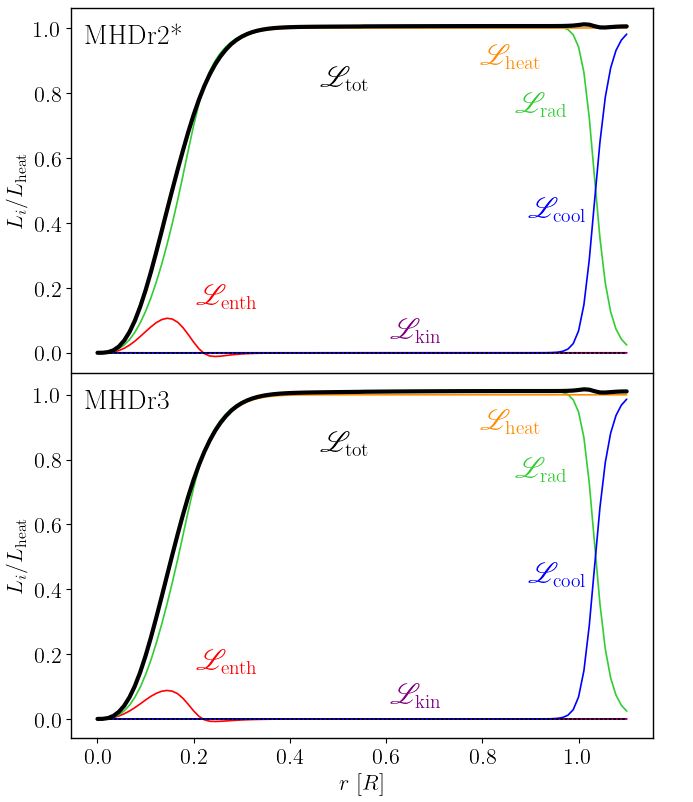}
    \caption{Luminosity contributions from kinetic energy, enthalpy,
      radiative, cooling and heating fluxes from MHDr2* (\textit{top
        panel}) and MHDr3 (\textit{bottom panel}).}
    \label{fig:lum}
\end{figure}

The contributions to the total luminosity from runs MHDr3 and MHDr2*
are shown in Figure~\ref{fig:lum}. $\mathscr{L}_\mathrm{visc}$ is not shown
because it is
negligible, reaching maximally $0.02\%$ of the total flux at $r
\approx 0.22 R$. The convective luminosity
$\mathscr{L}_\mathrm{conv} = \mathscr{L}_\mathrm{kin} +
\mathscr{L}_\mathrm{enth}$ is dominated by a positive
$\mathscr{L}_\mathrm{enth}$ in the convection zone ($0 < r \leq 0.2
R$). The kinetic energy flux is negligible which is expected in
the rapid rotation regime; see, for example
\cite{2024A&A...683A.221K}. In the majority of the star,
the contribution due to the radiative energy flux dominates. This is
due to the internal structure of the star, where the main energy
transfer mechanism is radiation (the radiative zone is roughly $80\%$
of the radial extent). Near the surface of the star the
cooling becomes effective and it dominates the total
luminosity. The rest of the runs from groups MHD and MHD* have
almost identical luminosity profiles. These profiles
are similar to those of \cite{Augustson-2016} and \cite{Brun-2005},
see their Figs.~3 and 13, respectively. The contribution of
$\mathscr{L}_\mathrm{enth}$ in the core is about $10 \%$ which is
somewhat smaller than in \cite{Brun-2005} ($\approx 25\%$), and in
\cite{Augustson-2016} ($\approx 40\%$).
In the zMHD group the luminosity profiles are very similar to
those shown in Figure~\ref{fig:lum}, with the obvious
differences that these profiles extend only to $r = 0.55 R$
and the cooling becomes effective around $r = 0.5R$.

\subsection{Rotational scaling of magnetic cycles \label{cycles}}

\begin{table*}[t!]
\centering
\caption{Comparison of values of $\beta$.}
\begin{tabular}{lcccc}
\hline\hline\noalign{\smallskip}
Data & Sample & $\beta$ & References\\
\hline\noalign{\smallskip}
\multirow{6}{*}{\begin{tabular}{@{}c@{}}Late-type stars \\ observations \end{tabular}}& Active branch & $0.48$ & \multirow{2}{*}{\cite{Brandenburg-1998} }\\
& Inactive branch & $0.46$ & \\
& S branch & -0.43 &  \cite{Saar-1999} \\
& 45 FGK stars & $<0$ & \cite{Boro-2018} \\
& 15 M dwarfs & $-1.02 \pm 0.06$ & \multirow{2}{*}{\cite{Irving-2023}} \\
& 40 FGK stars & $-0.81 \pm 0.17$ & \\
\hline
\multirow{8}{*}{\begin{tabular}{@{}c@{}}Solar-like stars \\ simulations \end{tabular}} & Slow rotators & $-0.73$ & \multirow{2}{*}{\cite{viviani-2018}} \\
& Fast rotators & $-0.50$ &  \\
 &  Global fit & $-0.99 \pm 0.05$ & \cite{Warnecke-2018} \\
& Global fit  & $-1.58 \pm 0.11$ & \cite{Strugarek-2018} \\
& Slow rotators & $-0.47 \pm 0.15$ & \multirow{2}{*}{\cite{Guerrero-2019}} \\
& Fast rotators & $1.17 \pm 0.05$ & \\
& Surface cycles & $-0.03 \pm 0.12$ & \multirow{2}{*}{\cite{Kapyla-2202} }\\
& Deep cycles & $-0.11 \pm 0.17$ &  \\
\hline
M dwarf simulations & Global fit &  $-1.30 \pm 0.26$ & \cite{ortizrodriguez-2023} \\
\hline
\multirow{3}{*}{ \begin{tabular}{@{}c@{}}A-type star \\ simulations\end{tabular}} & MHD group & $-1.22 \pm 0.05$ & \multirow{3}{*}{Present work} \\
& MHD* group & $-1.13 \pm 0.11$ & \\
& zMHD group & $-0.83 \pm 0.16$ &\\
\hline
\end{tabular}
\label{table-4}
\tablefoot{The data correspond to chromospheric observations and numerical simulations.}
\end{table*}

\begin{figure*}[t!]
    \centering
    \includegraphics[scale=0.5]{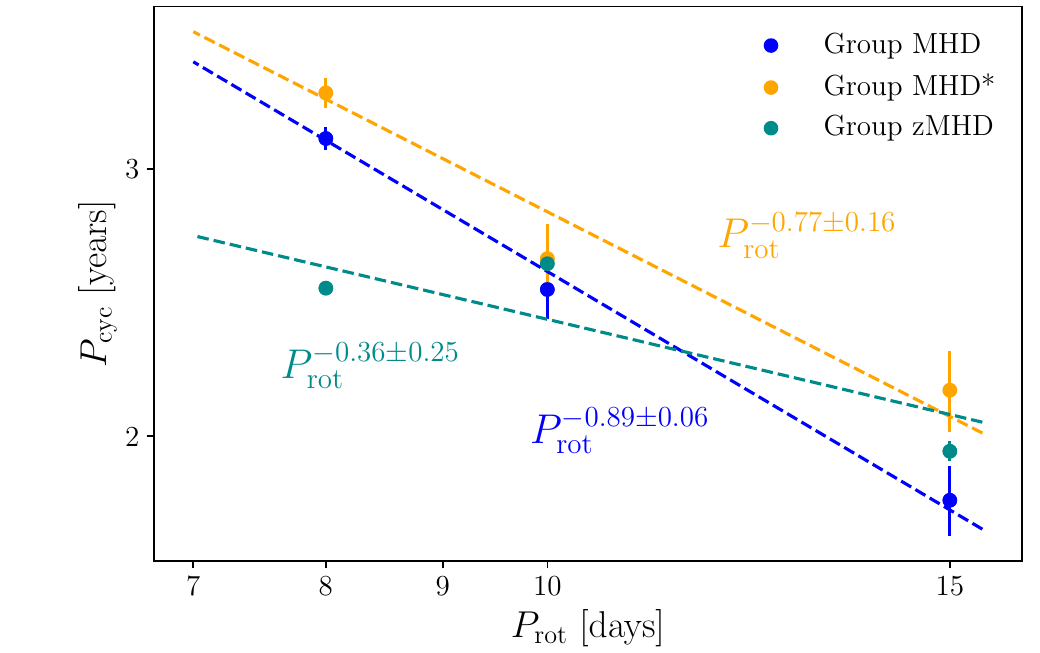}
    \includegraphics[scale=0.5]{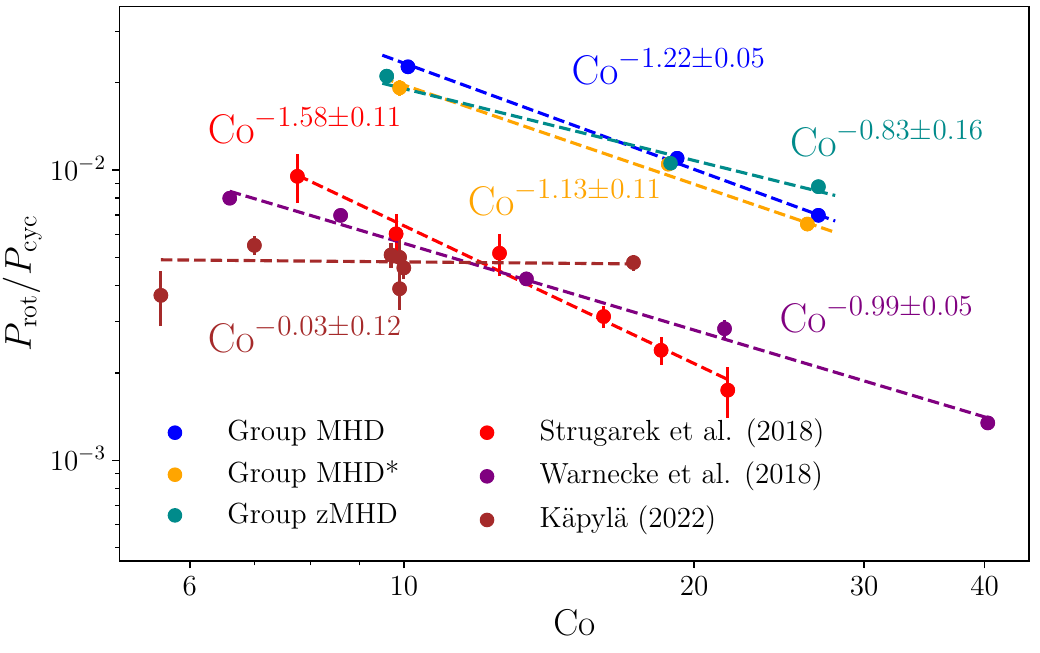}
    \caption{\textit{Left panel:} The magnetic cycle period
      $P_\mathrm{cyc}$ as a function of the rotation period
      $P_\mathrm{rot}$. \textit{Right panel:} Ratios of rotation to
      cycle period as a function of the Coriolis number. Additional
      data are from \citet[red]{Strugarek-2018},
      \citet[purple]{Warnecke-2018} and
      \citet[brown]{Kapyla-2202}. The dashed lines are the best
      power-law fits to the data. }
    \label{fig:Prot}
\end{figure*}

As discussed in Section~\ref{core-dynamos}, the averaged toroidal
magnetic fields in the core show clear cycles. To estimate
the magnetic cycle period $P_\mathrm{cyc}$
of the core dynamo we computed the spectral density of the magnetic
field at different latitudes from the active hemisphere using 
Welch's method \citep{Welch-1967}. The peak value of the power
spectrum corresponds to the frequency of the magnetic cycle, and its
inverse gives the cycle period. The mean value of $P_\mathrm{cyc}$
obtained from the considered latitudes with the respective standard
error is listed in the last column of Table~\ref{table-1}. In the full
star models there is a clear trend that $P_\mathrm{cyc}$ increases
with decreasing $P_\mathrm{rot}$. This is visible in left panel of
Figure~\ref{fig:Prot}. Furthermore, our best power-law fit shows
$P_\mathrm{cyc} \propto P_\mathrm{rot}^{-0.89  \pm  0.06}$ in the MHD
group, and $P_\mathrm{cyc} \propto P_\mathrm{rot}^{-0.77 \pm 0.16}$ in
the MHD* group. The relation between the magnetic cycle of the core
dynamo and rotation has not been studied before. However, simulations
of solar-like stars have shown similar results. Data from
\cite{Warnecke-2018}
indicate 
$P_\mathrm{cyc} \propto P_\mathrm{rot}^{-1.06}$, which is a relation
steeper than those found in our simulations. In the zMHD group the
magnetic cycle period seems to be less sensitive to rotation.
Our best fit shows $P_\mathrm{cyc} \propto
P_\mathrm{rot}^{-0.36 \pm 0.25}$, which is less steep than in the full
star models. This relation is similar to the $P_\mathrm{cyc} \propto
P_\mathrm{rot}^{-0.33 \pm 0.05}$ reported in \cite{Strugarek-2017}.

Observations of chromospheric activity suggest that the stellar magnetic
activity cycles are distributed in active 
and
inactive branches \citep[see e.g.][]{Brandenburg-1998,
Bohm-Vitense-2007}. These studies suggest a relation 
$P_\mathrm{rot}/P_\mathrm{cyc} \propto
\mathrm{Co}^{\beta}$ with $\beta > 0$.
However, the exact nature of these branches and 
the relation between stellar cycles and rotation are still under 
debate \citep[e.g.][]{Brandenburg-2017, Boro-2018, Bonanno-2022}. 
We present this relation in the core dynamo of an A-type star for the
first time, with $\beta \approx -1$ in all the sets. Our results are
shown in the right panel of Figure~\ref{fig:Prot}. Furthermore, our best
power law fits are
summarised in Table~\ref{table-4}. Data from other authors from 
observations and numerical simulations is also added for comparison 
purposes. 
From Table~\ref{table-4} it is visible that our
results are usually in agreement with solar-like simulations. However,
as discussed in Section~\ref{core-dynamos}, the expulsion of the
magnetic helicity has direct consequences in the cycle period, and
the mechanism responsible is expected to differ between core and
envelope dynamos. One possible explanation of why the scaling laws
seem to agree, is that all the solar-like simulations reported in
Table~\ref{core-dynamos} have been done at relatively modest magnetic 
Reynolds numbers, therefore, the direct effects of the magnetic 
helicity conservation might not be immediately visible.

\section{Summary and conclusions \label{conclusions}}

We present results from star-in-a-box simulations
\citep[see][]{Dobler-2006,
  Kapyla-2021} of a $2.2~M_\odot$ A-type star with a convective core of
roughly $20\%$ of the stellar radius.
We explore rotation periods from 8 to 20 days in three
sets of simulations. The full star sets MHD and MHD*
have somewhat different radial profiles of the kinematic viscosity
and the magnetic diffusivity. The third set (zMHD) consists of zoom-in
versions of the MHD group, where we model the convective core and a
part of the
radiative envelope at higher resolution. All of the core dynamos are
hemispheric and cyclic
with rms values of around 60 kG. 
Observational constrains of the internal magnetic field
  strength of early-type stars can be obtained via
  asteroseismology. \cite{Lecoanet-2022} estimated an upper limit of
  $B_r \approx 500~\mathrm{kG}$ at $r=0.18R$ for the B star HD 43317,
  based on the g-mode frequencies from
  \cite{Buysschaert-2018}. Our results are one order of magnitude
  lower than this (at $r=0.20R$ we find $\overline{B}_r^\mathrm{rms}
  \approx 20$ kG). However, in B-type stars the convective velocities
  in the core are expected to be roughly ten times higher than in
  A-type stars \citep{Browning-2004}. Therefore we would expect
  correspondingly stronger magnetic fields in the former; see, for
  example \cite{Augustson-2016}.
The dynamos in our simulations
remain hemispheric throughout the simulation time, except in
MHDr2* where the magnetic activity migrated from the southern
to the northern hemisphere. Similar behaviour was reported by
\cite{brown-2020} in fully convective M drawfs but the reason behind
it is not clear. Some magnetic fields reach the surface of the star in
our simulations. However, these fields are very weak everywhere except
at the poles, where the maximum rms value 
averaged over the poles ($-90^\circ > \theta > -75^\circ$,
and $75^\circ < \theta < 90^\circ$)
is in run MHDr3 with $\sim 0.47$
kG. In the group MHD* the surface magnetic fields are significantly
weaker due to the different radial profiles of diffusivities,
with maximum values
around $10^{-4}$ kG. These weak surface magnetic fields might be a
consequence
of the flows aligned with the rotation axis because of the
Taylor-Proudman theorem, due to
the high Coriolis numbers in our simulations. Such flows could
penetrate the radiative layers due to the unrealistically low
Richardson number there. 

All the simulations have approximately rigid rotation in a significant
part of their radiative envelope, and a solar-like differential
rotation in the convective core. On average, the core is rotating
slightly slower than the envelope.
Run MHDr1 has $\overline{\Omega}(r=0.2R)/\overline{\Omega}(r=R) 
\approx 0.8$, while the runs with dynamos have ratios very close to 
unity. Differential rotation has been observed in early-type stars via
asteroseismology \citep[see][and the references therein]
{2021osvm.confE..27B,2023Ap&SS.368..107B}. 
\cite{2014MNRAS.444..102K} reported the surface-to-core
rotation of the main-sequence A-type star KIC 11145123, finding that the
star is almost a rigid rotator, but the surface rotates slightly faster
than the core in agreement with our simulations. Furthermore,
most of the observed intermediate-mass main-sequence stars have nearly
rigid rotation, based on the average near-core and envelope rotation rates
\citep[see Fig.~4 of ][]{2023Ap&SS.368..107B}. 
In massive stars
($M >  9M_\odot$), the near-core rotation rate is typically
larger than that of the envelope. For example, recently,
\cite{Burssens-2023} deduced the core-to-surface radial rotation
profile of the B star HD 192575, finding that the convective core is
rotating between 1.4 and 6.3 times faster than the radiative envelope.

The hemispheric dynamos imprint an
asymmetry also on the rotation profile via magnetic quenching of the
the
differential rotation. There is no clear difference between groups in
their rotation profiles. As we increase the rotation rate, the
magnetic energy gets closer to the equipartition values with the
kinetic energy. The magnetic energy inside the core reaches $E_\mathrm{mag} \approx 0.4 E_\mathrm{kin}$ in the fast rotators MHDr4 and MHDr4*, while in the full star both energies are comparable. None of our runs
reached super-equipartition values like the fast rotators in \cite{Augustson-2016}. This might be a consequence of the
different values of the dimensionless diagnostic parameters in their
simulations, for example, the fluid Reynolds number \citep[computed as
  $\mathrm{Re}'/2\pi$ for a
proper comparison, see Appendix A of][]{Kapyla2017} are in the range
of 81-132, while those in our study range between 25 and 52. Furthermore,
$\mathrm{Re}_\mathrm{M}$ ($\mathrm{Rm'}/2\pi$) in their work ranges
from 324 to 490, being significantly higher than those in the
current study (17-36).

All groups have very similar
luminosity profiles, which are also similar to those in
\cite{Brun-2005} and \cite{Augustson-2016}. In general, the zoom models
show very similar results compared to the full star models. However,
in the zoom models the magnetic fields are slightly weaker (see
Tables~\ref{table-1} and \ref{table-energy}).
This is possibly a resolution effect.
Nevertheless, the changes are not drastic and the large-scale
structures are very similar giving us confidence in the robustness of
the results.

We find a relation $P_\mathrm{cyc} \propto P_\mathrm{rot}^\alpha$ with
$\alpha = -0.89 \pm 0.06$, $\alpha = -0.77 \pm 0.16$ and $\alpha =
-0.36\pm 0.25$ in the groups MHD, MHD* and zMHD,
respectively. Furthermore, we present a scaling of
$P_\mathrm{rot}/P_\mathrm{cyc} \propto \mathrm{Co}^\beta$, with $\beta
= -1.22 \pm 0.05$ in the MHD group, $\beta = -1.13 \pm 0.11$ in the
MHD* group and $\beta = -0.83 \pm 0.16$ in the zMHD group, being the
first time that this has been done with the core dynamo cycle of an
A-type star. Similar results with $\beta \approx -1$ were recently
reported by \cite{Irving-2023} and have been found in data from
simulations of other types of stars \citep[e.g.][]{Warnecke-2018,
  Strugarek-2018, viviani-2018, ortizrodriguez-2023}.
Furthermore, \cite{Bonanno-2022} reported a dichotomy in the
  relation $P_\mathrm{cyc} \propto P_\mathrm{rot}^\alpha$, in terms of
  $\omega_\mathrm{cyc}=2\pi/P_\mathrm{cyc}$ and $\Omega =
  2\pi/P_\mathrm{rot}$, where our results ($\alpha < 0$) agree with
  one of the reported branches (Group 2 in their Fig. 1). This branch
  is attributed to having older stars with higher metallicities than 
  the other. The reason of why our simulations fit with this group
  is currently unclear. However, there is
considerable debate regarding the scaling of stellar cycles as a
function of rotation both from observations as well as simulations
\citep[see, e.g.][and references therein]{2023SSRv..219...58K}.

Our simulations indicate that the magnetic field created by a core
dynamo is not enough to explain the large-scale structures observed at
the surface of Ap/Bp stars, in accordance with earlier
  analytical studies \citep[e.g.][]{MacGregor-2003}. This makes sense
considering the wide
range of magnetic field strengths observed at the surfaces of stars
whose convective cores are predicted to be of similar
size. Moreover, the surface magnetic fields of Ap/Bp stars
  often have simple geometries, for example, dipoles with a magnetic axis
  misaligned with the rotational axis, supporting the fossil field
  theory \citep[see][and the references
    therein]{2023Galax..11...40K}. Interesting, the internal magnetic
  field inferred by \cite{Lecoanet-2022} seems to be unlikely to reach
  exclusively with fossil fields, supporting the evidence of a strong
  core dynamo inside early-type stars. In future
studies, different mechanisms (for example, fossil fields, Tayler-Spruit
dynamo) should be included while the strong core dynamo is present. A
logical step to follow in the future is to add a fossil field into our
model, and to study how these magnetic configurations affect the nature
of the hemispheric dynamo that was obtained in all the magnetic
runs. Different initial configurations can be implemented as a way to
test their stability and how they interact with an existing core
dynamo, similarly to \cite{Featherstone-2009} but also modelling the
entire star and seeing how the resulting magnetic field behaves at the
stellar surface. Furthermore, more rotation rates can be explored,
although it is true that Ap stars can rotate very slowly
($P_\mathrm{rot} = 300$ years), most of them have rotation periods
between 1 and 10 days \citep{Braithwaite-2017}, so the “fast rotator
regime”, from 1 to 8 days, remains unexplored.

\begin{acknowledgements}
We thank the anonymous referee for providing useful comments on the 
manuscript.
JPH and DRGS gratefully acknowledge support by the ANID BASAL projects
ACE210002 and FB21003, as well as via Fondecyt Regular (project code
1201280) and via ANID Fondo 2022 QUIMAL 220002. The simulations were performed with resources provided by
the Kultrun Astronomy Hybrid Cluster via the projects Conicyt Quimal
\#170001, Conicyt PIA ACT172033, and Fondecyt Iniciacion 11170268.
PJK was supported in part by the Deutsche Forschungsgemeinschaft (DFG,
German Research Foundation) Heisenberg programme (grant No.\ KA
4825/4-1), and by the Munich Institute for Astro-, Particle and
BioPhysics (MIAPbP) which is funded by the DFG under Germany's
Excellence Strategy – EXC-2094 – 390783311. CAOR acknowledges financial support from ANID (DOCTORADO DAAD-BECAS CHILE/62220030) as well as financial support from DAAD (DAAD/BECAS Chile, 2023 - 57636841 ). FHN acknowledges funding from the program Unidad de Excelencia Mar\'ia Maeztu, reference CEX2020-001058-M.
\end{acknowledgements}

%
\bibliographystyle{aa}
\bibliography{astro.bib}
%

\end{document}